\documentstyle[12pt,aps,preprint,graphics,amsfonts]{revtex}

\newcommand{\q}{ {\bf q} }
\newcommand{\Q}{ {\bf Q} }
\renewcommand{\r}{{\it {\bf r}}}
\newcommand{\R}{{\it {\bf R}}}
\newcommand{\half}{\frac{1}{2} }

\begin{document}

\title{Ordering  at two length scales in comb-coil diblock copolymers 
consisting of only two different monomers}
\author{Rikkert J. Nap and Gerrit ten Brinke \\ 
{\it Department of Polymer Chemistry  and Materials Science Center,} \\
{\it University of Groningen, Nijenborgh 4, 9747 AG Groningen, The Netherlands}} 

\date{\today} 

\tighten

\maketitle 

{\abstract The microphase separated morphology of a melt of a specific
class of comb-coil diblock copolymers, consisting of an $AB$ comb
block and a linear homopolymer $A$ block, is analyzed in the weak
segregation limit. On increasing the length of the homopolymer $A$
block, the systems go through a characteristic series of structural
transitions. Starting from the pure comb copolymer the first series of
structures involve the short length scale followed by structures
involving the large length scale. A maximum of two critical points
exists. Furthermore, in the two parameter space, characterizing the
comb-coil diblock copolymer molecules considered, a non-trivial
bifurcation point exists beyond which the structure factor can have
two maxima (two correlation hole peaks).  }

\section{Introduction}

Diblock copolymer melts usually microphase separate with {\it one}
characteristic length. However, if more than two monomer types are
involved microphase separation frequently occurs at more than one
length scale
\cite{Breiner,Werner,Breiner2,Helsinki1,Helsinki2,Helsinki3,Ott}.  Several
examples, that are of direct interest for the present paper, can be
found in the experimental work of Ikkala and Ten Brinke and co-workers
\cite{Helsinki1,Helsinki2,Helsinki3}. There, comb-coil diblock
copolymers are investigated consisting of a poly(4-vinyl
pyridine)-{\it block}-polystyrene (P4VP-{\it b}-PS) diblock copolymer
with side chains (e.g. pentadecylphenol, PDP) attached by hydrogen
bonds to the P4VP-block. The resulting comb-coil diblock copolymers
show typical two length scale structure-inside-structure
morphologies. The PS-blocks microphase separate from the
P4VP(PDP)-blocks giving rise to the well known classical morphologies
depending on the volume fraction of either block. This structure
corresponds to the large length scale ordering and the order-disorder
transition temperature is, if present at all, very high. Inside the
P4VP(PDP) domains an additional short length scale lamellar ordering
takes place characterized by an easily accessible ( ca. $60^o$C)
order-disorder transition.

Still, two length scale ordering is not restricted to block copolymers
involving three or more chemically different monomers. To demonstrate
and analyze this, we studied in \cite{rikkert} the structure factor of
a block copolymer consisting of a linear homopolymer block of $A$
monomers linked to a comb copolymer block with a backbone of the same
kind of $A$ monomers and side chains consisting of $B$ monomers.
(An experimental example can be found in ref \cite{Beyer,Tsoukatos}.)
For convenience, to reduce the number of free parameters, the discussion
was restricted to side chains having a length equal to the backbone
length between two consecutive side chains.  Due to the architecture
of the molecule microphase separation can in principle occur at two
different length scales, either 'inside' the $AB$ comb block or
'between' the linear $A$ block and the $AB$ comb block. In the latter
case the behavior resembles that of a diblock copolymer, where one
block is the homopolymer $A$ block and the other block the $AB$ comb
block.

In \cite{rikkert} the analysis of the structure factor of this system
was presented and the main result was summarized in the form of a
so-called classification diagram. In this diagram, the horizontal and
vertical axes denote respectively the number of side chains and the
length of the homopolymer block; the latter expressed in units
corresponding to the side chain length. Together these two parameters
uniquely determine the molecular structure for the specific class of
comb-coil diblock copolymers considered, i.e., length of side chains
equal to length of backbone between consecutive side chains. The
position and the value of the absolute minimum of the inverse
structure factor determine the length scale of microphase separation
and the temperature at which the disordered phase becomes unstable
against structure formation, i.e., the spinodal. The classification
diagram presented showed which length scale was favored (absolute
minimum of the inverse structure factor) and whether the inverse
structure factor had one or two minima. One of the most striking
features was the existence of a non-trivial bifurcation point,
separating the region in the parameter space where the structure
factor has only one minimum from the region where the structure factor
can have two minima (two correlation hole peaks!).

In this paper we continue the analysis of the phase behavior of these
class of systems by calculating the free energy up to fourth order in
the weak segregation limit and constructing the corresponding phase
diagrams. In connection with this, also a brief discussion about the
phase behavior of pure comb copolymer melts is given, extending
results published in the past.

The paper is organized as follows. In section 2 the system parameters
are defined, in section 3 the theory is outlined and in section 4 the
results are discussed.
    
\section{Parameterization}

\begin{figure}
\scalebox{0.4}[0.4]{\rotatebox{-90}{
\includegraphics*{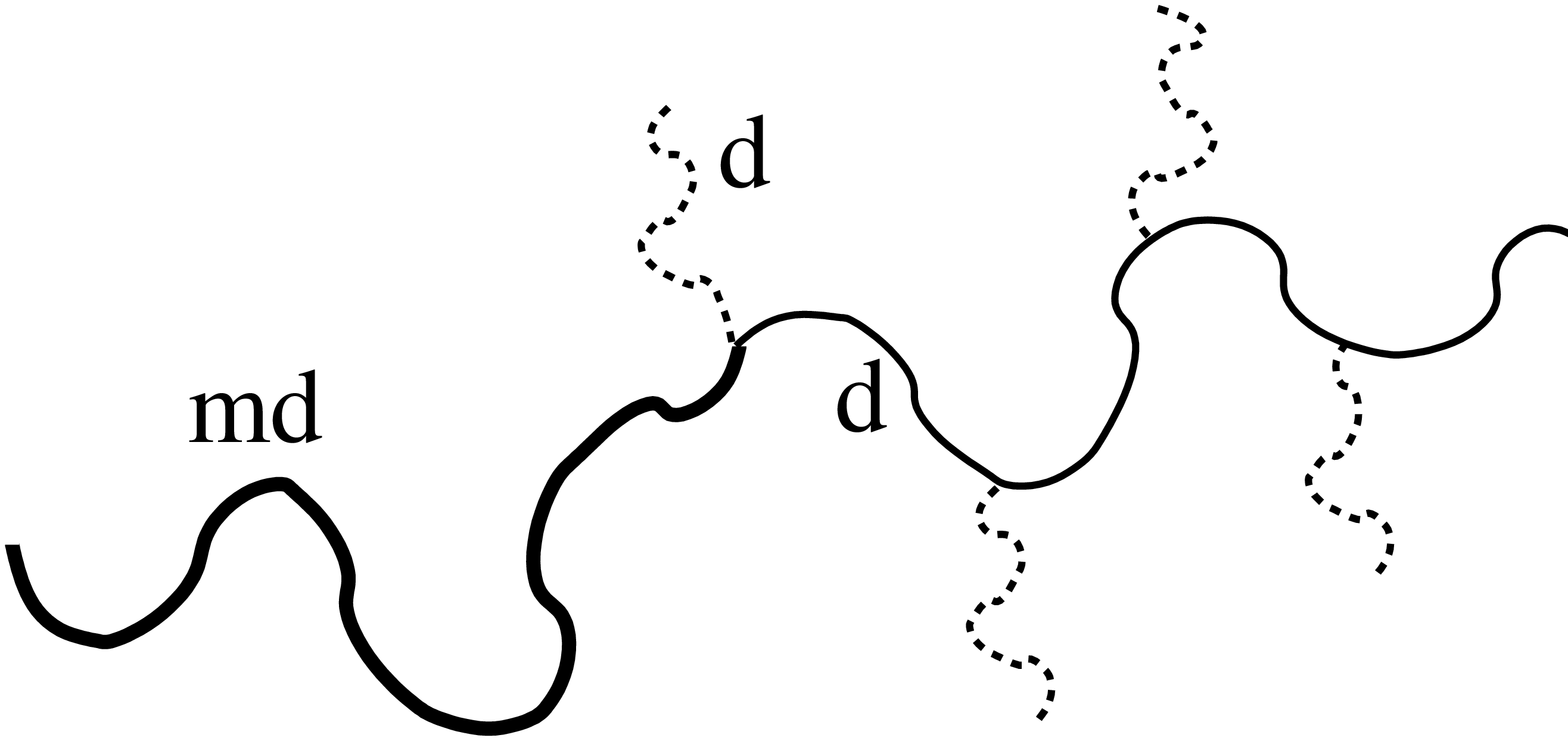}}}
\caption{Model of the comb-coil diblock copolymer molecule
studied. Note: for clarity the homopolymer block is indicated with a
thicker line than the backbone of the comb block. In this study both
are assumed to be chemically identical. The cartoon corresponds to:
number of branch points $n_t=4$, asymmetry parameter $t=0$ and one
side chain per branch point $\alpha=1$. 
The length of the side chains and length of the backbone between
two successive side chains equals $d$.
The length of the homopolymer block equals $md$, $m \in \mathbb{R}$.
}
\label{figI}
\end{figure}

We consider mono disperse polymer melts of comb copolymers and
comb-coil diblock copolymers.  The comb-coil diblock copolymer
molecules differ from the comb copolymer by the fact that an
additional $A$ homopolymer block is linked to the $AB$ comb copolymer
block, see Figure \ref{figI}.

First the parameterization of the comb block will be introduced. The
phase behavior of pure $AB$ comb copolymer melts, i.e., without the
$A$ homopolymer block, has already been discussed in several
papers.\cite{Balasz2,Balasz1,Benoit,Dobrynin}. However, all of these,
except for Dobrynin and Erukhimovich \cite{Dobrynin}, are restricted
to the calculation of the spinodals only.  Here, we adopt the notation
of \cite{Balasz2} and slightly extend it by incorporating the
possibility of having more than one side chain per branch point.

An $AB$ comb copolymer molecule consists of a backbone chain of
monomers of type $A$ to which side chains consisting of monomers of
type $B$ are attached.  The monomers (or segments) $A$ and $B$ are
assumed to be of equal size.  The number of backbone monomers equals
$N_A^b$ whereas the number of monomers per {\it one} side chain is
equal to $N_B$.  There are $n_t$ branch points which all have the same
functionality $z=\alpha + 2$. Here $\alpha$ is the number of side
chains linked to {\it one} branch point. Therefore, the number of side
chains equals $n_s=\alpha n_t$. The branch points are assumed to be
distributed regularly along the backbone chain. The number of backbone
segments between consecutive branch points is assumed to be equal to
$N_A^b/n_t$.  Obviously, the number of backbone segments before the
first branch point and after the last branch point must then also add up to
$N_A^b/n_t$.  Therefore, the number of backbone segments before the
first side chain and after the last side chain equals $t
N_A^b/n_t$ and $(1-t) N_A^b/n_t$ respectively, with $ t\in [0,1] $,
where $t$ is called the asymmetry parameter. \cite{Balasz2}

The $A$ homopolymer block has $N_A^h$ segments. Thus the total number
of $A$ segments in the comb-coil molecule is $N_A = N_A^b + N_A^h$ and that of
all segments is $N=N_A^b + N_A^h + n_s N_B $.  With the above
definitions the volume fractions are $f_A^b=N_A^b/N$, $f_A^h=N_A^h/N$
and $f_B=n_sN_B/N$.  Since the melt is assumed to be incompressible
$f_A^b + f_A^h + f_B=1$.

Together these parameters define a 6 dimensional compositional
parameter space: $(N_A^b,N_A^h,N_B,n_t,\alpha,t)$. However, in the
discussion of the comb-coil diblock copolymer melt we will restrict
this to a 3 dimensional subset by assuming $N_B=N_A^b/n_t=d$,
$\alpha=1$ and $t=0$, unless stated otherwise. Hence, the length of
the side chains is equal to the length of the backbone
chain between consecutive side chains and both will be denoted by
$d$. For the comb block this implies an equal amount of $A$ and $B$
segments; $f_A^b=f_B$.  Furthermore, we will also express $N_A^h$ in
terms of $d$ by defining $N_A^h=d m$, where $m\in \mathbb{R}^+$ is a
measure for the length of the homopolymer block in units of the side
chain length.  In terms of $m$ and $n_t$ the volume fraction of the
homopolymer block becomes $f_A^h= m /( 2 n_t +m ) $. Note that $N= (2
n_t + m) d $. The overall volume fraction of $A$ monomers is given by
$f_A=(m+n_t)/(m+2n_t)$. Because $N$ can be absorbed in the interaction
(Flory-Huggins) parameter $\chi$, we effectively reduced the number of
parameters to two, namely $(n_t,m)$ or $(n_t,f_A)$.

There is an important difference between our presentation of the
phase behavior of the pure comb copolymer melt and the comb-coil
diblock copolymer melt. In the former we will, for a given number of
branch points $n_t$, present the results as a function of the volume
fraction $f_A$. An increase in $f_A$ will correspond to a decrease in
the length of the side chains compared to the length of the backbone
between consecutive side chains. For the comb-coil diblock copolymers,
the architecture of the comb block is fixed (the length of the side
chain is always equal to the length of the backbone between
consecutive side chains), and the increase in $f_A$ (for given $n_t$)
is obtained by increasing the length $m$ of the homopolymer $A$-block.

\section{Theory}

The Hamiltonian of the copolymer melt is given by ($\beta = 1/kT$)
\cite{delaCruz,Edwards,Hong,Matsen0}.
\begin{equation}\label{I}
\beta H  = \beta ( H_0 + H_I ) =  \frac{3}{2 a^2} 
{\sum_{m,i}} \delta_c \int_0^{N_i}{\rm d}s \,
\left( \frac{\partial \R_i^m(s)}{\partial s} \right)^2 + 
\frac{1}{2} \sum_{\alpha \beta} \epsilon_{\alpha \beta}
\int_{V}{\rm d}^3 r \hat{\rho}_{\alpha}(\r)\hat{\rho}_{\beta}(\r). 
\end{equation}
The first part, $H_0$, the Edwards Hamiltonian \cite{Edwards},
accounts for the chain connectivity. It is a continuous model of a
Gaussian chain where a configuration of the {\it m}th molecule is
represented by a space curve, $\R^m(t)$. $a$ is the effective bond length.
 We subdivide $\R^m(t)$ the
total space curve in smaller space curves $\R_i^m(s)$. Here
$\R_i^m(s)$ corresponds to the part of the space curve $\R^m(t)$ which
is associated with the {\it i}th linear block in which the molecule
will be divided.  The variable $i$ enumerates the number of different
linear blocks.  The variable $s$ is a continuous variable whose value
ranges from $0$ to $N_i$ along the contour of the $i$th block; $N_i$
denotes the length of the $i$th block. We first enumerate the $A$
blocks ( $i \in {1,\cdots,n_t +2}$) and subsequently the $B$ blocks (
$i \in {n_t +3,n_s + n_t +3}$). Note that the $B$ blocks correspond to
the side chains and that the first $A$ block, corresponding to the
homopolymer block is not present in the case of pure comb
copolymers. Hence, $N_i$ is equal to
\begin{eqnarray} \label{II}
N_i &= & \left\{ 
\begin{array}{ccc}  
N_A^h            &  i=1 &  \mbox{homopolymer block}\\ 
\begin{array}{c}
t N_A^b / n_t    \\
N_A^b / n_t      \\
(1-t)N_A^b / n_t \\
\end{array} &
\begin{array}{c}
i=2                       \\
3 \leq i \leq n_t +1      \\
i= n_t +2                 \\
\end{array} &
\left.
\begin{array}{c}
\\
\\
\end{array}  \right\}  \mbox{backbone of comb block}\\
N_B  &  n_t+3 \leq i \leq n_s + n_t +3 & \mbox{side chains}\\
\end{array}
\right.
\end{eqnarray}
The configuration, i.e., the whole space curve $\R^m(t)$ is obtained by
linking the sub-chains $\R_i^m(s)$ together.  This translates into the
following constraint on $\R_i^m(s)$, represented with the function
$\delta_c$ in eq \ref{I} \cite{delaCruz} .
\begin{equation}\label{III}
\delta_c = \prod_{i=1}^{n_t+1} \delta(\R_i^m(N_i) -\R_{i+1}^m(0))  
\prod_{j=1,i>1}^\alpha\delta(\R_i^m(N_i) -\R_{i+j+nt+1}^m(0)).
\end{equation}
 
The second part, $H_I$, corresponds to the interaction
Hamiltonian. The $\epsilon_{\alpha \beta}$ are the dimensionless
effective interaction parameters between monomers of type $\alpha$ and
$\beta$, where $\alpha$ and $\beta$ denote either A or B.
$\hat{\rho}_{\alpha}(\r)$ is the microscopic monomer concentration of
monomer type $\alpha$ at position $\r$ and is given by
\begin{equation}\label{IV}
\hat{\rho}_{\alpha}(\r) = \sum_{m,i}\sigma_{\alpha}^{i} 
\int_0^{N_i}{\rm d}s \, \delta(\r -\R_i^m(s)).
\end{equation}
The variable $\sigma_\alpha^i$ denotes whether the $i$th block is of 
monomer type $\alpha $.
\begin{eqnarray}\label{V}
\sigma^{i}_{\alpha} &= &\left\{ 
\begin{array}{rc} 
1 & \mbox{if block {\it i} is of type $\alpha$} \\
0 & \mbox{otherwise} 
\end{array}
\right. 
\end{eqnarray}
On imposing incompressibility, i.e., $\hat{\rho}_A(\r)
+\hat{\rho}_B(\r) =1 $, $H_I$ can be rewritten in the more familiar
form 
\begin{equation}\label{VI}
H_I= -\chi \int_{V}{\rm d}^3 r\hat{\Psi}^2(\r) + C ,
\end{equation} 
where $C$ is an unimportant constant, $\chi$ is the Flory-Huggins
parameter defined by $\chi= (\epsilon_{AB} -1/2(\epsilon_{AA}
+\epsilon_{BB}))\beta^{-1}$.  $\hat{\Psi}$ represents the
concentration profile and equals the deviation of the microscopic
concentration of monomer $A$ from its average value; $\hat{\Psi}(\r) =
\hat{\rho}_A(\r) - f $.

The Landau mean-field free energy corresponds to an expansion in powers 
of the concentration profile $\psi(\q)=\langle \hat{\Psi}(\q)\rangle $:
\cite{Leibler,Vilgis}
\begin{equation}\label{IX}
F[\psi]= \sum_{n=2}^{4} \frac{1}{n! V^n} \sum_{\q_1 \cdots \q_n}
\Gamma_n (\q_1, \cdots, \q_n) \psi(\q_1)\cdots \psi(\q_n).
\end{equation}
The coefficients of the free energy expansion, the vertex functions
$\Gamma_n$ depend on the chemical single chain correlation functions;
$g_{\alpha_1\cdots\alpha_n}(\r_1,\cdots,\r_n)$.
\cite{Leibler,Vilgis,FML,Slot,Mayes} For a more detailed discussion on the
computational details of the correlations functions we refer to the
Appendix. Through minimizing eq \ref{IX} with respect 
to the concentration profile $\psi(\q)$ the 
equilibrium free energy is obtained. 

The concentration profile $\psi(\r)$ is expanded in a set of wave
functions which obey the symmetry of a given (periodic) structure.
\begin{equation}\label{XIII}
\psi(\r) = \sum_{m=1}^{\infty} \frac{A_m}{\sqrt{n_m}} \sum_{\Q \in H_m}
e^{i( {\bf Q} \r + \phi_{\bf Q})} \quad \psi(\q) = \sum_{m=1}^{\infty} 
\frac{A_m V}{\sqrt{n_m}}\sum_{\Q \in H_m}
e^{i \phi_\Q} \delta(\Q -\q)  
\end{equation}
where $H_m$ is a set of wave vectors $\Q$ which describes the symmetry
of the structure. $A_m$ and $\phi_\Q$ correspond to the amplitudes and
phases of the concentration profile. $m$ denotes the number of
harmonics or shells, $H_m$, and $n_m$ is half the number of vectors of
the $m$th harmonic.  Vectors belonging to the same harmonic have,
besides the same length, also the same amplitude.  The length of the
vectors in the first harmonic is denoted by $q_o$, Then the length of
the vectors in the higher harmonics are fixed because they are
multiples of $q_o$ ( depending on the symmetry of the phase).

The free energy expansion eq \ref{IX} is only valid in the weak
segregation limit (WSL), i.e., close to the critical point and
$|\psi|\ll 1$.  In the WSL the free energy is dominated by the wave
vector $q^*$, where $q^*$ is equal to the value of $q$ for which
$\Gamma_2(q)$ ascertains its minimum. Thus in the WSL we can set
$q_o$ equal to $q^*$. 

In this paper we will restrict ourselves to the {\it first harmonic
approximation} (FHA)\cite{Leibler} of the concentration
profile. Because of this, only the classical structures, i.e., 
disordered, lamellar, hexagonal and spherical (bcc) can be
considered.  The inclusion of higher harmonics is required to
discuss more complex structures such as the gyroid structure.
\cite{Matsen0,delaCruz2,Jones,Hamley1,Erukhimovich,Milner2,Aksimentiev,Alexander}
Because of the complicated nature of our structure factor, including
the possibility of two maxima, we decided to consider at first only
the FHA.

Two important quantities are the spinodal and the critical point.  The
spinodal denotes the line of instability of the disordered phase, the
line in phase space where the disordered phase ($\psi =0$) becomes
absolutely unstable against fluctuations.  It is given by the absolute
minimum of $\Gamma_2(q)$
\begin{equation}\label{XVI}
\left.\frac{\partial \Gamma_2(q)}{\partial q}\right|_{q^*}=0 \quad \mbox{and} \quad
\left.\frac{\partial^2 \Gamma_2(q)}{\partial q^2}\right|_{q^*} > 0 \quad \mbox{and} \quad \Gamma_2(q^*)=0    
\end{equation}
Note that the $\Gamma_2(q)$ is inversely proportional to the
structure factor. \cite{delaCruz}

The critical point is given by 
\begin{equation}\label{XVII}
spinodal  \wedge \quad \Gamma_3(q^\prime,q^{\prime},q^{\prime\prime})=0. \quad |q^\prime|=|q^{\prime}|=|q^{\prime\prime}|=|q^*|
\end{equation}
It is a point on the spinodal curve where also $\Gamma_3$ becomes
zero.  At the critical point the phase transition is continuous, i.e.,
with zero amplitude. Therefore close to the critical point the
amplitude will be small, thereby assuring the validity of the WSL.

\section{Discussion}

\subsection{Spinodals}

We start our discussion with a brief excursion to the pure comb
copolymer melt. The spinodals will be presented for fixed values of
$n_t$ in the usual way as a plot of $\chi N /n_t $ versus the volume
fraction $f_A$. Here, the total number of monomers of the comb
copolymer $N$ has been divided by $n_t$ in order to correctly account
for the length of the repeat unit. \cite{Balasz2,Benoit}. It is
important to realize again that in this case an increase in $f_A$
corresponds to a decrease in the side chain length compared to the
length of the backbone between two successive side chains. This is
different from the comb-coil diblock copolymers to be considered next,
where the volume fractions of the $A$ and $B$ monomers of the
comb-block have been fixed at equal value. In Figures \ref{figIII} and
\ref{figIV} the spinodals of the pure comb copolymer melts are
presented for $t=\half$, $\alpha=1$ (one side chain per branch point)
\cite{Balasz2} and $t=\half$, $\alpha=2$ (two side chains per branch
point), respectively.  On increasing $n_t$ the spinodal curves for the
different values of $n_t$ approach a limiting curve, as a comb
copolymer melt for sufficiently many side chains essentially
microphase separates on the length scale of {\it its} repeat unit. The
spinodal curves significantly differ from each other only for small
values of $n_t$, i.e., small number of side chains. The results
presented in Figure \ref{figIV} correspond to the case of two side
chains per branch point. Compared to one side chain per branch point
(Figure \ref{figIII}) the lowest point of the limiting spinodal
(i.e., $n_t \gg 1$) is considerably higher. This is due to the fact
that the entropic penalty associated with the formation of ordered
structures is larger for more complex molecules.

Next we turn to the comb-coil diblock copolymer melts, concentrating
on the specific class of systems for which the length of the side
chains equals the length of the backbone in between two successive
side chains. In our previous article \cite{rikkert}, the spinodals of
this class of comb-coil diblock copolymers were already investigated
in some detail. We start the discussion here by recalling the main
results, presenting them in a slightly different manner.  Figure
\ref{figV} shows the spinodal surface of $(\chi N)_s$ versus $n_t$ and
$m$ corresponding to the absolute minimum of $\Gamma_2(q)$.  The
change of $(\chi N)_s$ as a function of $m$ for a fixed value of $n_t$
is in line with the change in length scale of the microphase
separation, which changes from short to long for increasing values of
the homopolymer block size $m$. This fact is most clearly demonstrated
by Figure \ref{figVI}, where  $y=a^2 q^2 d /6$, which is 
a measure for the inverse length scale, has been
plotted as a function of $n_t$ and $m$.

A more detailed analysis \cite{rikkert} shows that $\Gamma_2(q)$ can
have either one minimum, which in turn might correspond to either the
short or the long length scale, or two minima corresponding to the two
length scales. In the latter case either the short or the long length
scale corresponds to the absolute minimum of $\Gamma_2(q)$ or,
potentially even more interesting, both minima have the same value.
In Figures \ref{figV} and \ref{figVI} the heavy lines drawn on the
surfaces indicate the boundaries between the region where
$\Gamma_2(q)$ has two minima and the region where $\Gamma_2(q)$ has
one minimum. The line where the values of both minima are equal is
also indicated.  The most striking observation is the existence of a
bifurcation point in the ($n_t,m$) parameter space.  For $n_t < 10$ or
$m < 4.8$ microphase separation is only possible at one length scale,
i.e., for parameter values satisfying either inequality $\Gamma_2(q)$
has only a single minimum.  Above the bifurcation point microphase
separation at two different length scales is possible, e.g. for $n_t >
10$ values of $m$ can be selected for which microphase separation
occurs at either length scale. There is a range of $(n_t,m)$ values
for which $\Gamma_2(q)$ has two minima.  Figure \ref{figVII} is a
projection of the above mentioned boundary lines onto the $(n_t,m)$
plane summarizing this behavior.
  
Above we noticed that a pure comb copolymer melt essentially
microphase separates on the length scale of {\it its} repeat
unit. This fact explains why the spinodal surface of Figure
\ref{figV}, corresponding to the short length scale (microphase
separation within the comb copolymer block; Fig. 2a), increases
linearly with $n_t$. On the other hand, the spinodal surface
corresponding to the long length scale (large value of $m$) remains
essentially 'constant' with changing $n_t$. The number of monomers
involved in the short length scale is $N/n_t$, that for the long
length scale is $N$.

\subsection{ Critical points } 

We consider the specific class of comb-coil diblock copolymers,
characterized by side chains of a length equal to the length of the
backbone between consecutive side chains. Therefore, an increase in
$f_A$ corresponds to an increase in the length of the homopolymer
block. Because of this, critical points will only be found for special
values of $n_t$ and $m$ ($f_A$). In Figure \ref{figVII} these critical
points are presented by the dashed and dotted line. The curve has two
branches, an upper and a lower branch. The upper branch is associated
with the large length scale structure, i.e., the diblock scale
separation. The lower branch corresponds to the short length scale,
i.e., separation within the comb block. The part of the lower branch
located inside region III, indicated by the dotted line, does not
correspond to true critical points anymore. In region III the absolute
minimum corresponds to the large length scale. The critical points
correspond to the relative minimum of $\Gamma_2$. For the position of
the absolute minimum of $\Gamma_2$, $\Gamma_3$ does not become zero.

For $n_t$ values ranging from 3 to 15, critical points are present for
two values of $m$. For $n_t>15$, the critical point corresponding to
the short length scale ordering no longer corresponds to the absolute
minimum of $\Gamma_2(q)$. In contrast, for $n_t=2$ there is no
critical point at all.  Of course, this is simply due to the specific
choice of parameters.  We selected $N_B=N_A^b/n_t=d$, $\alpha=1$ and
$t=0$ and for $n_t=2$ the critical point is located outside this
particular subset of parameters.

In \ref{figXII}, the critical points of the comb-coil diblock
copolymer melt are presented once more. This time in the $(n_t,f_A)$
plane. This presentation is better suited to show the upper branch
corresponding to the critical points of the large length scale
structure formation. It clearly illustrates the presence of two
critical points (two values of $m$) for $n_t>2$. Of course, for
$n_t>15$, the lower branch corresponds again to "pseudo" critical
points. In the same figure the critical points of three different comb
copolymer systems, characterized by $(\alpha,t)=(1,\half), (1,0)$ and$
(2,\half)$, are presented. In the case of pure comb copolymers, there is
obviously only one critical point for any given value of the number of
branch points $n_t$. Upon increasing $n_t$, the critical points
approach limiting values for the same reason as discussed before. For
a given number of side chains per branch, the effect of the asymmetry
parameter vanishes for large $n_t$. As already noticed before, the
effect of having two side chains per branch point ($\alpha=2$),
compared to one side chain per branch point ($\alpha=1$), is a much
more symmetric phase diagram. In the former case, the volume fraction
$f_A$ at the critical point approaches $0.497$ for increasing $n_t$. This
seems to be related to the fact that for $f_A=\half$, the repeat unit 
corresponds to a four arm star molecule.  Two arms consist of $A$ monomers, 
two arms (the side chains) consist of $B$ monomers. For $f_A=\half$ all four
arms have equal length and the repeat unit is completely symmetric.

\subsection{ Phase diagrams } 

Before we consider the comb-coil diblock copolymers, we again start
with the phase diagrams for the pure comb copolymer melts.
For two different comb copolymers,  
corresponding to $(\alpha,n_t,t)=(1,30,\half)$ and
$(2,30,\half)$, these are plotted in Figures \ref{figVIII} and
\ref{figIX}. The lines delineate the boundaries between the regions
where the different classical structures are stable.  The phase
boundaries are asymmetric and elevated compared to a diblock copolymer
melt. Increasing the number of side chains per branch point from one
to two raises the phase boundaries even more. However, the asymmetry
is strongly reduced. Altogether, the phase diagrams show the 'normal'
shape as expected from the literature on melts of complex block
copolymers \cite{Alexander,delaCruz,Mayes}.  The asymmetry is obvious
due to the fact that the architecture of a comb copolymer is not
invariant under the interchange of $A$ with $B$ or $f$ with $1-f$.
The elevation of the phase boundaries agrees with the general
observation that in more complex polymer melts the disordered phase is
more stable. As mentioned before, this is due to the fact that more
complex polymers will 'loose' relatively more entropy on structure
formation.

As became clear from the spinodal analysis, the phase behavior of
comb-coil diblock copolymers will of course be much richer. One way of
presenting a phase diagram for this class of systems is to fix $n_t$
and vary $m$. In this way we start with a pure comb copolymer $m=0$
and end with nearly pure homopolymer $m\gg1$. However, the existence
of a bifurcation point in parameter space separating two distinct
regions complicates matters. If we start with $n_t > 10$, the second
order vertex function will develop two minima on increasing $m$. First
the absolute minimum will correspond to the short length scale
ordering, then the two minima will attain the same value for some
specific value of $m$, then the absolute minimum will correspond to
the large length scale and, finally, the minimum belonging to the
short length scale will gradually disappear. As presented in the
Theory section, the analysis of the stability of various ordered
structures in block copolymer melts is usually based on a free energy
expansion using a single dominant wave vector $q^*$, where $q^*$
coincides with the $q$-value for which the second order vertex
function attains its minimum. So, it is quite obvious that the
analysis of the microphase separated morphology in the case where the
second order vertex function has two minima of nearly equal value
requires a completely different analysis involving the $q$-values of
both minima. (See ref \cite{Semenov}.)
Here we will restrict ourselves to the region in the
classification diagram left of the bifurcation point where there is
always only one minimum and the theory (WSL) outlined before can be
applied without any restriction.

The phase diagrams that are presented correspond to a fixed number
of branch points ($n_t$) and varying length of the homopolymer block
($m$), i.e., to slices in the $(n_t, m)$ plane for fixed $n_t$.  Figure
\ref{figX} and \ref{figXI} correspond to $n_t=2$ and $n_t=4$
respectively.  The phase diagrams are presented as $\chi N$ versus
$m$.

For $n_t=4$ the phase diagram has two critical points. The critical
point on the left signals the change from ordering on the short (comb)
length scale to ordering on the long (diblock) length scale, Figure
\ref{figVI}.  The critical point on the right corresponds to the
critical point of the effective diblock copolymer.  Throughout the
phase diagram the lamellar (resp. hexagonal and spherical) structure
is indicated with the same character L (resp. H and B.) It should be
realized, however, that the same symmetry is accompanied with strongly
different periodicities in different regions of the phase diagram.

To illustrate this we will walk through the phase diagram, Figure 12,
as a function of the length $m$ of the homopolymer block for a fixed
value of $\chi N = 60$, respectively $\chi N = 55$. The corresponding 
changes in length scale can be found from Fig. \ref{figVI}.

Let us start with $\chi N = 60$. For $m$ sufficiently small there is a
short length scale lamellar ordering, the layers being alternately
rich in side chain $B$-monomers and $A$-monomers. Then a short length
scale hexagonal structure is found with cylinders rich in side chain
$B$-monomers. Then a kind of in between "intermediate" length scale
lamellar structure is formed, followed by a large length scale
hexagonal structure with cylinders rich in homopolymer
$A$-monomers. Next, the sequence of structures follows the usual
pattern from lamellar (layers alternately rich in homopolymer block
and comb copolymer block) to hexagonal (cylinders rich in comb
copolymer block) to bcc (spheres rich in comb copolymer block) to
disordered.

For $\chi N = 55$, we start again with a short length scale lamellar
structure (layers alternately rich in side chain $B$-monomers and $A$
monomers) followed with a short length scale hexagonal structure
(cylinders rich in side chain $B$-monomers). However, then a sequence of
short length scale bcc (spheres rich in $B$-monomers), disordered state
and large length scale bcc (spheres rich in comb copolymer block) is
found. Next, the structures follow the same order as discussed above
for $\chi N = 60$.

For $n_t=2$ the phase diagram does not have any critical point. As
discussed before, this is simply due to the choice of parameters.
Going through the phase diagram at constant value of $\chi N$, e.g. $40$,
there is a much more gradual change in length scale. 

\section{Concluding Remark}

Phase diagram \ref{figXI} seems quite characteristic for the class of
comb-coil diblock copolymers considered. For sufficiently large values
of $\chi N$ a sequence of structures is found as a function of the
homopolymer block length $m$. The first part involves the short length
scale followed by large length scale ordered structures. This phase
diagram was calculated for values of $(n_t,m,)$ to the left of the
bifurcation point in the classification diagram (Figure
\ref{figVII}). To the left of the bifurcation point there is gradual
change from the short to the large length scale as a function of $m$.
To the right of the bifurcation point a similar change from short to
long length scale ordering will take place.  However, on increasing
$m$ we also traverse through the region where the second order vertex
function has two minima. On crossing the point where the minima are
equal a discontinuous change in length scale occurs. 
Here, for some values of $n_t$ and $m$ the occurrence of stable
structures which give rise to two minima should be expected, e.g. HML,
HPL or still more complex. The free energy expansion should involve
both wave vectors and this is the main topic of our current research
efforts.

\section{Acknowledgement} 
The authors are grateful to A.N. Morozov, Dr. M. Reenders, Dr
H.J. Angerman, Prof. Dr. S. Kuchanov and Prof. Dr. I. Erukhimovich for
helpful discussions.

\begin{appendix}

\section{Correlation functions}\label{correlators}

In this appendix we present in more detail the calculation of the
correlation functions. The n-point chemical correlation function
$g_{\alpha_1\cdots\alpha_n}(\r_1,\cdots,\r_n)$ denotes the probability
that at point $\r_i$ a monomer of type $\alpha_i $ is present for
$1\leq i \leq n $.
\begin{eqnarray}\label{C0}
g_{\alpha_1\cdots \alpha_n}(\r_1,\cdots,\r_n) = \frac{1}{N}
{\langle  \hat{\rho}^s_{\alpha_1}(\r_1) \cdots \hat{\rho}^s_{\alpha_n}(\r_n) \rangle }_0    \\ \label{C00}
{\langle \ldots \rangle}_0 = {\mathcal N} 
\int \prod_{mi} {\mathcal D} \R_i^m  \ldots e^{-{\hat H}_0([ \R_i^m]) }
\quad  {\langle 1 \rangle}_0 = 1 
\end{eqnarray}
In eq \ref{C0} $\hat{\rho}^s_\alpha$ denotes the concentration of a
single molecule. Consequently, the probability measure of the ensemble
average eq \ref{C00} is now also over one molecule only.
Defining $\hat{\rho}^s_\alpha(\r)=\sum_{i} \sigma_\alpha^i
\hat{\rho}^s_{i}(\r)$, where $\hat{\rho}^s_i(\r)$ is the microscopic
single molecule concentration of block $i$ of at position $\r$.  Eq
\ref{C0} can be rewritten as
\begin{equation}\label{CI}
g_{\alpha_1\cdots \alpha_n}(\r_1,\cdots,\r_n)  = 
\frac{1}{N} \sum_{ i_1 , \cdots , i_n } \sigma_\alpha^{i_1} 
\cdots \sigma_\alpha^{i_n} 
{\langle  
\hat{\rho}^s_{i_1}(\r_1) \cdots \hat{\rho}^s_{i_n}(\r_n) 
\rangle }_0 
\end{equation}
${\langle \hat{\rho}^s_{i_1}(\r_1) \cdots \hat{\rho}^s_{i_n}(\r_n)
\rangle }_0 $ has a similar meaning as ${\langle
\hat{\rho}^s_{\alpha_1}(\r_1) \cdots \hat{\rho}^s_{\alpha_n}(\r_n)
\rangle }_0 $; it is the probability of finding a monomers of block
$i$ at position $\r_i$, for $1\leq i \leq n $.
We omit the factor $1/N$ in the remaining discussion and write $G = N
g$.

As the generic example we take the number of side chains equal to two
($n_t=2$).  Schematically the molecule is pictured below
\begin{equation}\label{CII}
\setlength{\unitlength}{4144sp}%
\begingroup\makeatletter\ifx\SetFigFont\undefined%
\gdef\SetFigFont#1#2#3#4#5{%
  \reset@font\fontsize{#1}{#2pt}%
  \fontfamily{#3}\fontseries{#4}\fontshape{#5}%
  \selectfont}%
\fi\endgroup%
\begin{picture}(2384,967)(474,-241)
\thinlines
\put(1261,-61){\circle*{90}}
\put(2071,-61){\circle*{90}}
\thicklines
\special{ps: gsave 0 0 0 setrgbcolor}\put(856,-61){\line( 1, 0){360}}
\special{ps: grestore}\special{ps: gsave 0 0 0 setrgbcolor}\put(1306,-61){\line( 1, 0){360}}
\special{ps: grestore}\special{ps: gsave 0 0 0 setrgbcolor}\put(496,-61){\line( 1, 0){360}}
\special{ps: grestore}\special{ps: gsave 0 0 0 setrgbcolor}\put(1666,-61){\line( 1, 0){360}}
\special{ps: grestore}\special{ps: gsave 0 0 0 setrgbcolor}\put(2116,-61){\line( 1, 0){360}}
\special{ps: grestore}\special{ps: gsave 0 0 0 setrgbcolor}\put(2476,-61){\line( 1, 0){360}}
\special{ps: grestore}\special{ps: gsave 0 0 0 setrgbcolor}\multiput(1261,-16)(0.00000,80.00000){5}{\line( 0, 1){ 40.000}}
\special{ps: grestore}\special{ps: gsave 0 0 0 setrgbcolor}\multiput(2071,-16)(0.00000,80.00000){5}{\line( 0, 1){ 40.000}}
\special{ps: grestore}\special{ps: gsave 0 0 0 setrgbcolor}\multiput(2071,344)(0.00000,80.00000){5}{\line( 0, 1){ 40.000}}
\special{ps: grestore}\special{ps: gsave 0 0 0 setrgbcolor}\multiput(1261,344)(0.00000,80.00000){5}{\line( 0, 1){ 40.000}}
\special{ps: grestore}\put(856,-241){\makebox(0,0)[lb]{\smash{\SetFigFont{12}{14.4}{\rmdefault}{\mddefault}{\updefault}\special{ps: gsave 0 0 0 setrgbcolor}1\special{ps: grestore}}}}
\put(1306,299){\makebox(0,0)[lb]{\smash{\SetFigFont{12}{14.4}{\rmdefault}{\mddefault}{\updefault}\special{ps: gsave 0 0 0 setrgbcolor}4\special{ps: grestore}}}}
\put(2116,299){\makebox(0,0)[lb]{\smash{\SetFigFont{12}{14.4}{\rmdefault}{\mddefault}{\updefault}\special{ps: gsave 0 0 0 setrgbcolor}5\special{ps: grestore}}}}
\put(1621,-241){\makebox(0,0)[lb]{\smash{\SetFigFont{12}{14.4}{\rmdefault}{\mddefault}{\updefault}\special{ps: gsave 0 0 0 setrgbcolor}2\special{ps: grestore}}}}
\put(2431,-241){\makebox(0,0)[lb]{\smash{\SetFigFont{12}{14.4}{\rmdefault}{\mddefault}{\updefault}\special{ps: gsave 0 0 0 setrgbcolor}3\special{ps: grestore}}}}
\end{picture}
\end{equation}
Here $N_1=N_2=N_3=N_A$ and $N_4=N_5=N_B$.
Now we write $G_{AB}$ in terms of $G_{ij}$, the block correlation 
functions. 
\begin{eqnarray}
\lefteqn{ G_{AB}(\q_1,\q_2) = G_{14}(\q_1,\q_2) + G_{15}(\q_1,\q_2) +} \quad \nonumber  \\
& &  G_{25}(\q_1,\q_2) +G_{24}(\q_1,\q_2) + G_{34}(\q_1,\q_2) + G_{35}(\q_1,\q_2) \label{CIII}
\end{eqnarray}
These functions can, with the above picture in mind, be represented
graphically as follows.

\begin{eqnarray}\label{CIV}
\setlength{\unitlength}{4144sp}%
\begingroup\makeatletter\ifx\SetFigFont\undefined%
\gdef\SetFigFont#1#2#3#4#5{%
  \reset@font\fontsize{#1}{#2pt}%
  \fontfamily{#3}\fontseries{#4}\fontshape{#5}%
  \selectfont}%
\fi\endgroup%
\begin{picture}(4418,2362)(173,-1636)
\put(1126,119){\makebox(0,0)[lb]{\smash{\SetFigFont{12}{14.4}{\rmdefault}{\mddefault}{\updefault}\special{ps: gsave 0 0 0 setrgbcolor}4\special{ps: grestore}}}}
\thicklines
\special{ps: gsave 0 0 0 setrgbcolor}\put(586,-241){\line(-1, 0){360}}
\special{ps: grestore}\special{ps: gsave 0 0 0 setrgbcolor}\put(1036,-241){\line(-1, 0){360}}
\special{ps: grestore}\special{ps: gsave 0 0 0 setrgbcolor}\multiput(1081,344)(0.00000,80.00000){5}{\line( 0, 1){ 40.000}}
\special{ps: grestore}\special{ps: gsave 0 0 0 setrgbcolor}\multiput(1081,164)(0.00000,-80.00000){5}{\line( 0,-1){ 40.000}}
\special{ps: grestore}\thinlines
\put(1081,-241){\circle*{90}}
\thicklines
\special{ps: gsave 0 0 0 setrgbcolor}\put(3826,-241){\line(-1, 0){360}}
\special{ps: grestore}\special{ps: gsave 0 0 0 setrgbcolor}\put(4276,-241){\line(-1, 0){360}}
\special{ps: grestore}\special{ps: gsave 0 0 0 setrgbcolor}\multiput(4321,344)(0.00000,80.00000){5}{\line( 0, 1){ 40.000}}
\special{ps: grestore}\special{ps: gsave 0 0 0 setrgbcolor}\multiput(4321,164)(0.00000,-80.00000){5}{\line( 0,-1){ 40.000}}
\special{ps: grestore}\thinlines
\put(4321,-241){\circle*{90}}
\thicklines
\special{ps: gsave 0 0 0 setrgbcolor}\put(721,-1456){\line( 1, 0){360}}
\special{ps: grestore}\special{ps: gsave 0 0 0 setrgbcolor}\put(271,-1456){\line( 1, 0){360}}
\special{ps: grestore}\special{ps: gsave 0 0 0 setrgbcolor}\multiput(226,-871)(0.00000,80.00000){5}{\line( 0, 1){ 40.000}}
\special{ps: grestore}\special{ps: gsave 0 0 0 setrgbcolor}\multiput(226,-1051)(0.00000,-80.00000){5}{\line( 0,-1){ 40.000}}
\special{ps: grestore}\thinlines
\put(226,-1456){\circle*{90}}
\thicklines
\special{ps: gsave 0 0 0 setrgbcolor}\put(4006,-1411){\line( 1, 0){360}}
\special{ps: grestore}\special{ps: gsave 0 0 0 setrgbcolor}\put(3556,-1411){\line( 1, 0){360}}
\special{ps: grestore}\special{ps: gsave 0 0 0 setrgbcolor}\multiput(3511,-826)(0.00000,80.00000){5}{\line( 0, 1){ 40.000}}
\special{ps: grestore}\special{ps: gsave 0 0 0 setrgbcolor}\multiput(3511,-1006)(0.00000,-80.00000){5}{\line( 0,-1){ 40.000}}
\special{ps: grestore}\thinlines
\put(3511,-1411){\circle*{90}}
\put(2296,-241){\circle*{90}}
\thicklines
\special{ps: gsave 0 0 0 setrgbcolor}\put(2701,-241){\line(-1, 0){360}}
\special{ps: grestore}\special{ps: gsave 0 0 0 setrgbcolor}\put(1801,-241){\line(-1, 0){360}}
\special{ps: grestore}\special{ps: gsave 0 0 0 setrgbcolor}\put(2251,-241){\line(-1, 0){360}}
\special{ps: grestore}\special{ps: gsave 0 0 0 setrgbcolor}\multiput(3106,164)(0.00000,-80.00000){5}{\line( 0,-1){ 40.000}}
\special{ps: grestore}\special{ps: gsave 0 0 0 setrgbcolor}\multiput(3106,344)(0.00000,80.00000){5}{\line( 0, 1){ 40.000}}
\special{ps: grestore}\special{ps: gsave 0 0 0 setrgbcolor}\put(3061,-241){\line(-1, 0){360}}
\special{ps: grestore}\thinlines
\put(3106,-241){\circle*{90}}
\put(2206,-1456){\circle*{90}}
\thicklines
\special{ps: gsave 0 0 0 setrgbcolor}\put(1801,-1456){\line( 1, 0){360}}
\special{ps: grestore}\special{ps: gsave 0 0 0 setrgbcolor}\put(2701,-1456){\line( 1, 0){360}}
\special{ps: grestore}\special{ps: gsave 0 0 0 setrgbcolor}\put(2251,-1456){\line( 1, 0){360}}
\special{ps: grestore}\special{ps: gsave 0 0 0 setrgbcolor}\multiput(1396,-1051)(0.00000,-80.00000){5}{\line( 0,-1){ 40.000}}
\special{ps: grestore}\special{ps: gsave 0 0 0 setrgbcolor}\multiput(1396,-871)(0.00000,80.00000){5}{\line( 0, 1){ 40.000}}
\special{ps: grestore}\special{ps: gsave 0 0 0 setrgbcolor}\put(1441,-1456){\line( 1, 0){360}}
\special{ps: grestore}\thinlines
\put(1396,-1456){\circle*{90}}
\put(631,-421){\makebox(0,0)[lb]{\smash{\SetFigFont{12}{14.4}{\rmdefault}{\mddefault}{\updefault}\special{ps: gsave 0 0 0 setrgbcolor}1\special{ps: grestore}}}}
\put(1261,209){\makebox(0,0)[lb]{\smash{\SetFigFont{12}{14.4}{\rmdefault}{\mddefault}{\updefault}\special{ps: gsave 0 0 0 setrgbcolor}+\special{ps: grestore}}}}
\put(1846,-421){\makebox(0,0)[lb]{\smash{\SetFigFont{12}{14.4}{\rmdefault}{\mddefault}{\updefault}\special{ps: gsave 0 0 0 setrgbcolor}1\special{ps: grestore}}}}
\put(3151,119){\makebox(0,0)[lb]{\smash{\SetFigFont{12}{14.4}{\rmdefault}{\mddefault}{\updefault}\special{ps: gsave 0 0 0 setrgbcolor}5\special{ps: grestore}}}}
\put(4366, 74){\makebox(0,0)[lb]{\smash{\SetFigFont{12}{14.4}{\rmdefault}{\mddefault}{\updefault}\special{ps: gsave 0 0 0 setrgbcolor}5\special{ps: grestore}}}}
\put(3916,-421){\makebox(0,0)[lb]{\smash{\SetFigFont{12}{14.4}{\rmdefault}{\mddefault}{\updefault}\special{ps: gsave 0 0 0 setrgbcolor}2\special{ps: grestore}}}}
\put(4591,209){\makebox(0,0)[lb]{\smash{\SetFigFont{12}{14.4}{\rmdefault}{\mddefault}{\updefault}\special{ps: gsave 0 0 0 setrgbcolor}+\special{ps: grestore}}}}
\put(721,-1636){\makebox(0,0)[lb]{\smash{\SetFigFont{12}{14.4}{\rmdefault}{\mddefault}{\updefault}\special{ps: gsave 0 0 0 setrgbcolor}2\special{ps: grestore}}}}
\put(271,-1141){\makebox(0,0)[lb]{\smash{\SetFigFont{12}{14.4}{\rmdefault}{\mddefault}{\updefault}\special{ps: gsave 0 0 0 setrgbcolor}4\special{ps: grestore}}}}
\put(1441,-1096){\makebox(0,0)[lb]{\smash{\SetFigFont{12}{14.4}{\rmdefault}{\mddefault}{\updefault}\special{ps: gsave 0 0 0 setrgbcolor}4\special{ps: grestore}}}}
\put(2566,-421){\makebox(0,0)[lb]{\smash{\SetFigFont{12}{14.4}{\rmdefault}{\mddefault}{\updefault}\special{ps: gsave 0 0 0 setrgbcolor}2\special{ps: grestore}}}}
\put(2701,-1636){\makebox(0,0)[lb]{\smash{\SetFigFont{12}{14.4}{\rmdefault}{\mddefault}{\updefault}\special{ps: gsave 0 0 0 setrgbcolor}3\special{ps: grestore}}}}
\put(1801,-1636){\makebox(0,0)[lb]{\smash{\SetFigFont{12}{14.4}{\rmdefault}{\mddefault}{\updefault}\special{ps: gsave 0 0 0 setrgbcolor}2\special{ps: grestore}}}}
\put(4276,-1591){\makebox(0,0)[lb]{\smash{\SetFigFont{12}{14.4}{\rmdefault}{\mddefault}{\updefault}\special{ps: gsave 0 0 0 setrgbcolor}3\special{ps: grestore}}}}
\put(1126,-1006){\makebox(0,0)[lb]{\smash{\SetFigFont{12}{14.4}{\rmdefault}{\mddefault}{\updefault}\special{ps: gsave 0 0 0 setrgbcolor}+\special{ps: grestore}}}}
\put(181,209){\makebox(0,0)[lb]{\smash{\SetFigFont{12}{14.4}{\rmdefault}{\mddefault}{\updefault}\special{ps: gsave 0 0 0 setrgbcolor}=\special{ps: grestore}}}}
\put(3601,-1051){\makebox(0,0)[lb]{\smash{\SetFigFont{12}{14.4}{\rmdefault}{\mddefault}{\updefault}\special{ps: gsave 0 0 0 setrgbcolor}5\special{ps: grestore}}}}
\put(3106,-961){\makebox(0,0)[lb]{\smash{\SetFigFont{12}{14.4}{\rmdefault}{\mddefault}{\updefault}\special{ps: gsave 0 0 0 setrgbcolor}+\special{ps: grestore}}}}
\put(3376,164){\makebox(0,0)[lb]{\smash{\SetFigFont{12}{14.4}{\rmdefault}{\mddefault}{\updefault}\special{ps: gsave 0 0 0 setrgbcolor}+\special{ps: grestore}}}}
\put(1081,254){\circle{90}}
\put(631,-241){\circle{90}}
\put(4321,254){\circle{90}}
\put(3871,-241){\circle{90}}
\put(226,-961){\circle{90}}
\put(676,-1456){\circle{90}}
\put(3511,-916){\circle{90}}
\put(3961,-1411){\circle{90}}
\put(3106,254){\circle{90}}
\put(1846,-241){\circle{90}}
\put(1396,-961){\circle{90}}
\put(2656,-1456){\circle{90}}
\end{picture}
\end{eqnarray}
The open dot represents the q-vector ``flowing'' through the block,
the solid dot denotes a branch point.  The solid lines correspond to
$A$ blocks and the dashed lines to $B$ blocks.

The calculation of $G_{25}$ and $G_{15}$ amounts to performing an
integration over the monomers in the block, i.e., along the lines.
\begin{eqnarray}\label{CV}
G_{25}=\int_0^{N_A} \int_0^{N_B}{\rm d} i {\rm d} j e^{-x( i+ j)} \\ \label{CVI}
G_{15}= e^{-x N_A} G_{25} 
\end{eqnarray}
Here $x=a^2 q^2/6$ with $q=|\q_1|=|\q_2|$. Note that the delta
function $\delta(\q_1 + \q_2)$ for momentum conservation, which arises
naturally when performing the integration of eq \ref{CI} is not
written down explicitly. When adding $G_{25}$ and $G_{15}$ we obtain
$G_{(12)5}$, the block correlation function of the combined block of
blocks 1,2 and $5$. It has the same form as $G_{25}$ only the $A$
block is twice as long.
\begin{equation}\label{CVII}
G_{(12)5}=G_{25} + G_{15} = \int_0^{2N_A} \int_0^{N_B}{\rm d} i {\rm d} j e^{-x( i+ j)} 
\end{equation}
Graphically this is represented as
\begin{equation}\label{CVIII}
G_{(12)5} 
\begin{array}{l}
\setlength{\unitlength}{4144sp}%
\begingroup\makeatletter\ifx\SetFigFont\undefined%
\gdef\SetFigFont#1#2#3#4#5{%
  \reset@font\fontsize{#1}{#2pt}%
  \fontfamily{#3}\fontseries{#4}\fontshape{#5}%
  \selectfont}%
\fi\endgroup%
\begin{picture}(5048,1192)(91,-421)
\thinlines
\put(5086,-196){\circle*{90}}
\put(4276,-196){\circle{90}}
\thicklines
\special{ps: gsave 0 0 0 setrgbcolor}\put(5041,-196){\line(-1, 0){360}}
\special{ps: grestore}\special{ps: gsave 0 0 0 setrgbcolor}\put(4681,-196){\line(-1, 0){360}}
\special{ps: grestore}\special{ps: gsave 0 0 0 setrgbcolor}\put(4231,-196){\line(-1, 0){360}}
\special{ps: grestore}\special{ps: gsave 0 0 0 setrgbcolor}\multiput(5086,209)(0.00000,-80.00000){5}{\line( 0,-1){ 40.000}}
\special{ps: grestore}\special{ps: gsave 0 0 0 setrgbcolor}\multiput(5086,389)(0.00000,80.00000){5}{\line( 0, 1){ 40.000}}
\special{ps: grestore}\special{ps: gsave 0 0 0 setrgbcolor}\put(3871,-196){\line(-1, 0){360}}
\special{ps: grestore}\put(4591,-376){\makebox(0,0)[lb]{\smash{\SetFigFont{12}{14.4}{\rmdefault}{\mddefault}{\updefault}\special{ps: gsave 0 0 0 setrgbcolor}2\special{ps: grestore}}}}
\put(5131,119){\makebox(0,0)[lb]{\smash{\SetFigFont{12}{14.4}{\rmdefault}{\mddefault}{\updefault}\special{ps: gsave 0 0 0 setrgbcolor}5\special{ps: grestore}}}}
\put(3871,-376){\makebox(0,0)[lb]{\smash{\SetFigFont{12}{14.4}{\rmdefault}{\mddefault}{\updefault}\special{ps: gsave 0 0 0 setrgbcolor}1\special{ps: grestore}}}}
\special{ps: gsave 0 0 0 setrgbcolor}\put(586,-241){\line(-1, 0){360}}
\special{ps: grestore}\special{ps: gsave 0 0 0 setrgbcolor}\put(1036,-241){\line(-1, 0){360}}
\special{ps: grestore}\special{ps: gsave 0 0 0 setrgbcolor}\multiput(1081,344)(0.00000,80.00000){5}{\line( 0, 1){ 40.000}}
\special{ps: grestore}\special{ps: gsave 0 0 0 setrgbcolor}\multiput(1081,164)(0.00000,-80.00000){5}{\line( 0,-1){ 40.000}}
\special{ps: grestore}\thinlines
\put(1081,-241){\circle*{90}}
\put(2296,-241){\circle*{90}}
\thicklines
\special{ps: gsave 0 0 0 setrgbcolor}\put(2701,-241){\line(-1, 0){360}}
\special{ps: grestore}\special{ps: gsave 0 0 0 setrgbcolor}\put(1801,-241){\line(-1, 0){360}}
\special{ps: grestore}\special{ps: gsave 0 0 0 setrgbcolor}\put(2251,-241){\line(-1, 0){360}}
\special{ps: grestore}\special{ps: gsave 0 0 0 setrgbcolor}\multiput(3106,164)(0.00000,-80.00000){5}{\line( 0,-1){ 40.000}}
\special{ps: grestore}\special{ps: gsave 0 0 0 setrgbcolor}\multiput(3106,344)(0.00000,80.00000){5}{\line( 0, 1){ 40.000}}
\special{ps: grestore}\special{ps: gsave 0 0 0 setrgbcolor}\put(3061,-241){\line(-1, 0){360}}
\special{ps: grestore}\thinlines
\put(3106,-241){\circle*{90}}
\put(1846,-421){\makebox(0,0)[lb]{\smash{\SetFigFont{12}{14.4}{\rmdefault}{\mddefault}{\updefault}\special{ps: gsave 0 0 0 setrgbcolor}1\special{ps: grestore}}}}
\put(3151,119){\makebox(0,0)[lb]{\smash{\SetFigFont{12}{14.4}{\rmdefault}{\mddefault}{\updefault}\special{ps: gsave 0 0 0 setrgbcolor}5\special{ps: grestore}}}}
\put(2566,-421){\makebox(0,0)[lb]{\smash{\SetFigFont{12}{14.4}{\rmdefault}{\mddefault}{\updefault}\special{ps: gsave 0 0 0 setrgbcolor}2\special{ps: grestore}}}}
\put(3376,164){\makebox(0,0)[lb]{\smash{\SetFigFont{12}{14.4}{\rmdefault}{\mddefault}{\updefault}\special{ps: gsave 0 0 0 setrgbcolor}=\special{ps: grestore}}}}
\put( 91,209){\makebox(0,0)[lb]{\smash{\SetFigFont{12}{14.4}{\rmdefault}{\mddefault}{\updefault}\special{ps: gsave 0 0 0 setrgbcolor}=\special{ps: grestore}}}}
\put(676,-421){\makebox(0,0)[lb]{\smash{\SetFigFont{12}{14.4}{\rmdefault}{\mddefault}{\updefault}\special{ps: gsave 0 0 0 setrgbcolor}2\special{ps: grestore}}}}
\put(1171, 74){\makebox(0,0)[lb]{\smash{\SetFigFont{12}{14.4}{\rmdefault}{\mddefault}{\updefault}\special{ps: gsave 0 0 0 setrgbcolor}5\special{ps: grestore}}}}
\put(1306,209){\makebox(0,0)[lb]{\smash{\SetFigFont{12}{14.4}{\rmdefault}{\mddefault}{\updefault}\special{ps: gsave 0 0 0 setrgbcolor}+\special{ps: grestore}}}}
\put(5086,299){\circle{90}}
\put(1081,254){\circle{90}}
\put(631,-241){\circle{90}}
\put(3106,254){\circle{90}}
\put(1846,-241){\circle{90}}
\end{picture}
\end{array}
\end{equation}
Similarly we contract $G_{24}$ and $G_{34}$ into $G_{(23)4}$ and so on.
Finally, all the different diagrams of eq \ref{CIII} can be contracted
into 'one' diagram.
\begin{equation}\label{CIX}
G_{AB}(q)= 2 \sum\limits_{J} 
\left( 
\begin{array}{l}
\setlength{\unitlength}{4144sp}%
\begingroup\makeatletter\ifx\SetFigFont\undefined%
\gdef\SetFigFont#1#2#3#4#5{%
  \reset@font\fontsize{#1}{#2pt}%
  \fontfamily{#3}\fontseries{#4}\fontshape{#5}%
  \selectfont}%
\fi\endgroup%
\begin{picture}(930,1192)(294,-646)
\thicklines
\special{ps: gsave 0 0 0 setrgbcolor}\put(1126,-421){\line(-1, 0){360}}
\special{ps: grestore}\special{ps: gsave 0 0 0 setrgbcolor}\multiput(1171,-16)(0.00000,-80.00000){5}{\line( 0,-1){ 40.000}}
\special{ps: grestore}\special{ps: gsave 0 0 0 setrgbcolor}\multiput(1171,164)(0.00000,80.00000){5}{\line( 0, 1){ 40.000}}
\special{ps: grestore}\thinlines
\put(1171,-421){\circle*{90}}
\thicklines
\special{ps: gsave 0 0 0 setrgbcolor}\put(676,-421){\line(-1, 0){360}}
\special{ps: grestore}\put(811,-286){\makebox(0,0)[lb]{\smash{\SetFigFont{12}{14.4}{\rmdefault}{\mddefault}{\updefault}\special{ps: gsave 0 0 0 setrgbcolor}q\special{ps: grestore}}}}
\put(1171,-646){\makebox(0,0)[lb]{\smash{\SetFigFont{12}{14.4}{\rmdefault}{\mddefault}{\updefault}\special{ps: gsave 0 0 0 setrgbcolor}J\special{ps: grestore}}}}
\thinlines
\put(1171, 74){\circle{90}}
\put(721,-421){\circle{90}}
\end{picture}  
\end{array} 
\right)  
\end{equation}
The length of the lines is not related anymore with its length; J
indicates its length.  The variable $J$ is the summation variable over
the different branch points.  Summation over branch points is equal to
summing over the different lengths of all the $A$ blocks in `front' of
the side chains, i.e., in this example $J=N_A$ or $2N_A$.  The
prefactor 2 arises from the fact that the molecule is symmetric thus
$G_{14}=G_{35}$ and $G_{14}=G_{34}$; the backbone ends are of equal
length.  However, when the ends of the backbone have different lengths
they are not equal: $N_2 \not= N_{n_t+2}$. So for the comb copolymer
parameterized as discussed in the theory section ($t\not=\half$),
there will be two distinct diagrams.  Note that the example molecule
has no asymmetry parameter.  Thus for a comb parameterized by
$(\alpha=1,n_t,t)$ the correlation function $G_{AB}$ is:
\begin{equation}\label{CX}
G_{AB}(q)=  \sum\limits_{J} 
\left( 
\begin{array}{l}
\setlength{\unitlength}{4144sp}%
\begingroup\makeatletter\ifx\SetFigFont\undefined%
\gdef\SetFigFont#1#2#3#4#5{%
  \reset@font\fontsize{#1}{#2pt}%
  \fontfamily{#3}\fontseries{#4}\fontshape{#5}%
  \selectfont}%
\fi\endgroup%
\begin{picture}(2429,1192)(294,-646)
\thicklines
\special{ps: gsave 0 0 0 setrgbcolor}\put(1126,-421){\line(-1, 0){360}}
\special{ps: grestore}\special{ps: gsave 0 0 0 setrgbcolor}\multiput(1171,-16)(0.00000,-80.00000){5}{\line( 0,-1){ 40.000}}
\special{ps: grestore}\special{ps: gsave 0 0 0 setrgbcolor}\multiput(1171,164)(0.00000,80.00000){5}{\line( 0, 1){ 40.000}}
\special{ps: grestore}\thinlines
\put(1171,-421){\circle*{90}}
\thicklines
\special{ps: gsave 0 0 0 setrgbcolor}\put(676,-421){\line(-1, 0){360}}
\special{ps: grestore}\put(811,-286){\makebox(0,0)[lb]{\smash{\SetFigFont{12}{14.4}{\rmdefault}{\mddefault}{\updefault}\special{ps: gsave 0 0 0 setrgbcolor}q\special{ps: grestore}}}}
\thinlines
\put(1846,-421){\circle*{90}}
\thicklines
\special{ps: gsave 0 0 0 setrgbcolor}\put(2341,-421){\line( 1, 0){360}}
\special{ps: grestore}\special{ps: gsave 0 0 0 setrgbcolor}\put(1891,-421){\line( 1, 0){360}}
\special{ps: grestore}\special{ps: gsave 0 0 0 setrgbcolor}\multiput(1846,-16)(0.00000,-80.00000){5}{\line( 0,-1){ 40.000}}
\special{ps: grestore}\special{ps: gsave 0 0 0 setrgbcolor}\multiput(1846,164)(0.00000,80.00000){5}{\line( 0, 1){ 40.000}}
\special{ps: grestore}\put(1171,-646){\makebox(0,0)[lb]{\smash{\SetFigFont{12}{14.4}{\rmdefault}{\mddefault}{\updefault}\special{ps: gsave 0 0 0 setrgbcolor}J\special{ps: grestore}}}}
\put(1486, 29){\makebox(0,0)[lb]{\smash{\SetFigFont{12}{14.4}{\rmdefault}{\mddefault}{\updefault}\special{ps: gsave 0 0 0 setrgbcolor}+\special{ps: grestore}}}}
\put(1801,-646){\makebox(0,0)[lb]{\smash{\SetFigFont{12}{14.4}{\rmdefault}{\mddefault}{\updefault}\special{ps: gsave 0 0 0 setrgbcolor}J\special{ps: grestore}}}}
\put(2116,-331){\makebox(0,0)[lb]{\smash{\SetFigFont{12}{14.4}{\rmdefault}{\mddefault}{\updefault}\special{ps: gsave 0 0 0 setrgbcolor}q\special{ps: grestore}}}}
\thinlines
\put(1171, 74){\circle{90}}
\put(721,-421){\circle{90}}
\put(2296,-421){\circle{90}}
\put(1846, 74){\circle{90}}
\end{picture}  
\end{array} 
\right)  
\end{equation}
Where $J$ runs over $\left\{t,t+N_A/n_t,t+2N_A/n_t,\cdots\right\}$.

So far this method may seems rather clumsy.  Because the integral and
subsequent summations can be done quite easily, see \cite{Balasz2}.
However, when we want to calculate third and fourth order correlation
functions this method becomes a useful book keeping tool and a concise
way of writing down correlation functions.  As an example we present
two third order correlation functions in their diagrammatic
representation.
\begin{eqnarray}
& G_{AAB}(q,k,p)= \nonumber \\
& \sum\limits_{J} 
\left( 
\begin{array}{l}
\setlength{\unitlength}{4144sp}%
\begingroup\makeatletter\ifx\SetFigFont\undefined%
\gdef\SetFigFont#1#2#3#4#5{%
  \reset@font\fontsize{#1}{#2pt}%
  \fontfamily{#3}\fontseries{#4}\fontshape{#5}%
  \selectfont}%
\fi\endgroup%
\begin{picture}(5174,1282)(159,-601)
\thicklines
\special{ps: gsave 0 0 0 setrgbcolor}\put(1441,-286){\line(-1, 0){360}}
\special{ps: grestore}\special{ps: gsave 0 0 0 setrgbcolor}\multiput(1486,119)(0.00000,-80.00000){5}{\line( 0,-1){ 40.000}}
\special{ps: grestore}\special{ps: gsave 0 0 0 setrgbcolor}\multiput(1486,299)(0.00000,80.00000){5}{\line( 0, 1){ 40.000}}
\special{ps: grestore}\thinlines
\put(1486,-286){\circle*{90}}
\thicklines
\special{ps: gsave 0 0 0 setrgbcolor}\put(541,-286){\line(-1, 0){360}}
\special{ps: grestore}\put(1306,164){\makebox(0,0)[lb]{\smash{\SetFigFont{12}{14.4}{\rmdefault}{\mddefault}{\updefault}\special{ps: gsave 0 0 0 setrgbcolor}p\special{ps: grestore}}}}
\put(451,-196){\makebox(0,0)[lb]{\smash{\SetFigFont{12}{14.4}{\rmdefault}{\mddefault}{\updefault}\special{ps: gsave 0 0 0 setrgbcolor}q\special{ps: grestore}}}}
\put(946,-196){\makebox(0,0)[lb]{\smash{\SetFigFont{12}{14.4}{\rmdefault}{\mddefault}{\updefault}\special{ps: gsave 0 0 0 setrgbcolor}k\special{ps: grestore}}}}
\special{ps: gsave 0 0 0 setrgbcolor}\put(991,-286){\line(-1, 0){360}}
\special{ps: grestore}\thinlines
\put(4006,-331){\circle*{90}}
\thicklines
\special{ps: gsave 0 0 0 setrgbcolor}\put(4501,-331){\line( 1, 0){360}}
\special{ps: grestore}\special{ps: gsave 0 0 0 setrgbcolor}\put(4951,-331){\line( 1, 0){360}}
\special{ps: grestore}\special{ps: gsave 0 0 0 setrgbcolor}\put(4051,-331){\line( 1, 0){360}}
\special{ps: grestore}\special{ps: gsave 0 0 0 setrgbcolor}\multiput(4006, 74)(0.00000,-80.00000){5}{\line( 0,-1){ 40.000}}
\special{ps: grestore}\special{ps: gsave 0 0 0 setrgbcolor}\multiput(4006,254)(0.00000,80.00000){5}{\line( 0, 1){ 40.000}}
\special{ps: grestore}\put(4096,164){\makebox(0,0)[lb]{\smash{\SetFigFont{12}{14.4}{\rmdefault}{\mddefault}{\updefault}\special{ps: gsave 0 0 0 setrgbcolor}p\special{ps: grestore}}}}
\put(4906,-241){\makebox(0,0)[lb]{\smash{\SetFigFont{12}{14.4}{\rmdefault}{\mddefault}{\updefault}\special{ps: gsave 0 0 0 setrgbcolor}q\special{ps: grestore}}}}
\put(4411,-196){\makebox(0,0)[lb]{\smash{\SetFigFont{12}{14.4}{\rmdefault}{\mddefault}{\updefault}\special{ps: gsave 0 0 0 setrgbcolor}k\special{ps: grestore}}}}
\put(3961,-556){\makebox(0,0)[lb]{\smash{\SetFigFont{12}{14.4}{\rmdefault}{\mddefault}{\updefault}\special{ps: gsave 0 0 0 setrgbcolor}J\special{ps: grestore}}}}
\thinlines
\put(2791,-331){\circle*{90}}
\thicklines
\special{ps: gsave 0 0 0 setrgbcolor}\put(3646,-331){\line(-1, 0){360}}
\special{ps: grestore}\special{ps: gsave 0 0 0 setrgbcolor}\put(3196,-331){\line(-1, 0){360}}
\special{ps: grestore}\special{ps: gsave 0 0 0 setrgbcolor}\multiput(2791, 74)(0.00000,-80.00000){5}{\line( 0,-1){ 40.000}}
\special{ps: grestore}\special{ps: gsave 0 0 0 setrgbcolor}\multiput(2791,254)(0.00000,80.00000){5}{\line( 0, 1){ 40.000}}
\special{ps: grestore}\special{ps: gsave 0 0 0 setrgbcolor}\put(2746,-331){\line(-1, 0){360}}
\special{ps: grestore}\special{ps: gsave 0 0 0 setrgbcolor}\put(2296,-331){\line(-1, 0){360}}
\special{ps: grestore}\put(2611,119){\makebox(0,0)[lb]{\smash{\SetFigFont{12}{14.4}{\rmdefault}{\mddefault}{\updefault}\special{ps: gsave 0 0 0 setrgbcolor}p\special{ps: grestore}}}}
\put(2341,-196){\makebox(0,0)[lb]{\smash{\SetFigFont{12}{14.4}{\rmdefault}{\mddefault}{\updefault}\special{ps: gsave 0 0 0 setrgbcolor}q\special{ps: grestore}}}}
\put(3196,-241){\makebox(0,0)[lb]{\smash{\SetFigFont{12}{14.4}{\rmdefault}{\mddefault}{\updefault}\special{ps: gsave 0 0 0 setrgbcolor}k\special{ps: grestore}}}}
\put(2746,-601){\makebox(0,0)[lb]{\smash{\SetFigFont{12}{14.4}{\rmdefault}{\mddefault}{\updefault}\special{ps: gsave 0 0 0 setrgbcolor}J\special{ps: grestore}}}}
\put(2071,164){\makebox(0,0)[lb]{\smash{\SetFigFont{12}{14.4}{\rmdefault}{\mddefault}{\updefault}\special{ps: gsave 0 0 0 setrgbcolor}+\special{ps: grestore}}}}
\put(3376,164){\makebox(0,0)[lb]{\smash{\SetFigFont{12}{14.4}{\rmdefault}{\mddefault}{\updefault}\special{ps: gsave 0 0 0 setrgbcolor}+\special{ps: grestore}}}}
\put(1441,-511){\makebox(0,0)[lb]{\smash{\SetFigFont{12}{14.4}{\rmdefault}{\mddefault}{\updefault}\special{ps: gsave 0 0 0 setrgbcolor}J\special{ps: grestore}}}}
\thinlines
\put(1486,209){\circle{90}}
\put(586,-286){\circle{90}}
\put(1036,-286){\circle{90}}
\put(4456,-331){\circle{90}}
\put(4906,-331){\circle{90}}
\put(4006,164){\circle{90}}
\put(3241,-331){\circle{90}}
\put(2791,164){\circle{90}}
\put(2341,-331){\circle{90}}
\end{picture}
\end{array} 
\right) + q \leftrightarrow k 
\end{eqnarray}
\begin{eqnarray}
& G_{ABB}(q,k,p)= \nonumber \\
& \sum\limits_{J<K} 
\left( 
\begin{array}{l}
\setlength{\unitlength}{4144sp}%
\begingroup\makeatletter\ifx\SetFigFont\undefined%
\gdef\SetFigFont#1#2#3#4#5{%
  \reset@font\fontsize{#1}{#2pt}%
  \fontfamily{#3}\fontseries{#4}\fontshape{#5}%
  \selectfont}%
\fi\endgroup%
\begin{picture}(5084,1192)(204,-511)
\thicklines
\special{ps: gsave 0 0 0 setrgbcolor}\multiput(3241,119)(0.00000,-80.00000){5}{\line( 0,-1){ 40.000}}
\special{ps: grestore}\special{ps: gsave 0 0 0 setrgbcolor}\multiput(3241,299)(0.00000,80.00000){5}{\line( 0, 1){ 40.000}}
\special{ps: grestore}\special{ps: gsave 0 0 0 setrgbcolor}\multiput(2341,119)(0.00000,-80.00000){5}{\line( 0,-1){ 40.000}}
\special{ps: grestore}\special{ps: gsave 0 0 0 setrgbcolor}\multiput(2341,299)(0.00000,80.00000){5}{\line( 0, 1){ 40.000}}
\special{ps: grestore}\special{ps: gsave 0 0 0 setrgbcolor}\put(3196,-286){\line(-1, 0){360}}
\special{ps: grestore}\special{ps: gsave 0 0 0 setrgbcolor}\put(2746,-286){\line(-1, 0){360}}
\special{ps: grestore}\thinlines
\put(2341,-286){\circle*{90}}
\put(3241,-286){\circle*{90}}
\thicklines
\special{ps: gsave 0 0 0 setrgbcolor}\multiput(1891,119)(0.00000,-80.00000){5}{\line( 0,-1){ 40.000}}
\special{ps: grestore}\special{ps: gsave 0 0 0 setrgbcolor}\multiput(1891,299)(0.00000,80.00000){5}{\line( 0, 1){ 40.000}}
\special{ps: grestore}\special{ps: gsave 0 0 0 setrgbcolor}\multiput(1081,119)(0.00000,-80.00000){5}{\line( 0,-1){ 40.000}}
\special{ps: grestore}\special{ps: gsave 0 0 0 setrgbcolor}\multiput(1081,299)(0.00000,80.00000){5}{\line( 0, 1){ 40.000}}
\special{ps: grestore}\thinlines
\put(1081,-286){\circle*{90}}
\thicklines
\special{ps: gsave 0 0 0 setrgbcolor}\put(1846,-286){\line(-1, 0){360}}
\special{ps: grestore}\special{ps: gsave 0 0 0 setrgbcolor}\put(1486,-286){\line(-1, 0){360}}
\special{ps: grestore}\special{ps: gsave 0 0 0 setrgbcolor}\put(1036,-286){\line(-1, 0){360}}
\special{ps: grestore}\special{ps: gsave 0 0 0 setrgbcolor}\put(586,-286){\line(-1, 0){360}}
\special{ps: grestore}\thinlines
\put(1891,-286){\circle*{90}}
\thicklines
\special{ps: gsave 0 0 0 setrgbcolor}\multiput(3601,119)(0.00000,-80.00000){5}{\line( 0,-1){ 40.000}}
\special{ps: grestore}\special{ps: gsave 0 0 0 setrgbcolor}\multiput(3601,299)(0.00000,80.00000){5}{\line( 0, 1){ 40.000}}
\special{ps: grestore}\special{ps: gsave 0 0 0 setrgbcolor}\multiput(4411,119)(0.00000,-80.00000){5}{\line( 0,-1){ 40.000}}
\special{ps: grestore}\special{ps: gsave 0 0 0 setrgbcolor}\multiput(4411,299)(0.00000,80.00000){5}{\line( 0, 1){ 40.000}}
\special{ps: grestore}\thinlines
\put(4411,-286){\circle*{90}}
\thicklines
\special{ps: gsave 0 0 0 setrgbcolor}\put(3646,-286){\line( 1, 0){360}}
\special{ps: grestore}\special{ps: gsave 0 0 0 setrgbcolor}\put(4006,-286){\line( 1, 0){360}}
\special{ps: grestore}\special{ps: gsave 0 0 0 setrgbcolor}\put(4456,-286){\line( 1, 0){360}}
\special{ps: grestore}\special{ps: gsave 0 0 0 setrgbcolor}\put(4906,-286){\line( 1, 0){360}}
\special{ps: grestore}\thinlines
\put(3601,-286){\circle*{90}}
\put(2071,164){\makebox(0,0)[lb]{\smash{\SetFigFont{12}{14.4}{\rmdefault}{\mddefault}{\updefault}\special{ps: gsave 0 0 0 setrgbcolor}+\special{ps: grestore}}}}
\put(3376,164){\makebox(0,0)[lb]{\smash{\SetFigFont{12}{14.4}{\rmdefault}{\mddefault}{\updefault}\special{ps: gsave 0 0 0 setrgbcolor}+\special{ps: grestore}}}}
\put(496,-151){\makebox(0,0)[lb]{\smash{\SetFigFont{12}{14.4}{\rmdefault}{\mddefault}{\updefault}\special{ps: gsave 0 0 0 setrgbcolor}q\special{ps: grestore}}}}
\put(2566,-196){\makebox(0,0)[lb]{\smash{\SetFigFont{12}{14.4}{\rmdefault}{\mddefault}{\updefault}\special{ps: gsave 0 0 0 setrgbcolor}q\special{ps: grestore}}}}
\put(4726,-196){\makebox(0,0)[lb]{\smash{\SetFigFont{12}{14.4}{\rmdefault}{\mddefault}{\updefault}\special{ps: gsave 0 0 0 setrgbcolor}q\special{ps: grestore}}}}
\put(856,164){\makebox(0,0)[lb]{\smash{\SetFigFont{12}{14.4}{\rmdefault}{\mddefault}{\updefault}\special{ps: gsave 0 0 0 setrgbcolor}k\special{ps: grestore}}}}
\put(2431,164){\makebox(0,0)[lb]{\smash{\SetFigFont{12}{14.4}{\rmdefault}{\mddefault}{\updefault}\special{ps: gsave 0 0 0 setrgbcolor}k\special{ps: grestore}}}}
\put(1711,164){\makebox(0,0)[lb]{\smash{\SetFigFont{12}{14.4}{\rmdefault}{\mddefault}{\updefault}\special{ps: gsave 0 0 0 setrgbcolor}p\special{ps: grestore}}}}
\put(3061,164){\makebox(0,0)[lb]{\smash{\SetFigFont{12}{14.4}{\rmdefault}{\mddefault}{\updefault}\special{ps: gsave 0 0 0 setrgbcolor}p\special{ps: grestore}}}}
\put(4231,164){\makebox(0,0)[lb]{\smash{\SetFigFont{12}{14.4}{\rmdefault}{\mddefault}{\updefault}\special{ps: gsave 0 0 0 setrgbcolor}p\special{ps: grestore}}}}
\put(3691,164){\makebox(0,0)[lb]{\smash{\SetFigFont{12}{14.4}{\rmdefault}{\mddefault}{\updefault}\special{ps: gsave 0 0 0 setrgbcolor}k\special{ps: grestore}}}}
\put(1036,-511){\makebox(0,0)[lb]{\smash{\SetFigFont{12}{14.4}{\rmdefault}{\mddefault}{\updefault}\special{ps: gsave 0 0 0 setrgbcolor}J\special{ps: grestore}}}}
\put(1846,-511){\makebox(0,0)[lb]{\smash{\SetFigFont{12}{14.4}{\rmdefault}{\mddefault}{\updefault}\special{ps: gsave 0 0 0 setrgbcolor}K\special{ps: grestore}}}}
\put(2296,-511){\makebox(0,0)[lb]{\smash{\SetFigFont{12}{14.4}{\rmdefault}{\mddefault}{\updefault}\special{ps: gsave 0 0 0 setrgbcolor}J\special{ps: grestore}}}}
\put(3601,-511){\makebox(0,0)[lb]{\smash{\SetFigFont{12}{14.4}{\rmdefault}{\mddefault}{\updefault}\special{ps: gsave 0 0 0 setrgbcolor}J\special{ps: grestore}}}}
\put(3196,-511){\makebox(0,0)[lb]{\smash{\SetFigFont{12}{14.4}{\rmdefault}{\mddefault}{\updefault}\special{ps: gsave 0 0 0 setrgbcolor}K\special{ps: grestore}}}}
\put(4366,-511){\makebox(0,0)[lb]{\smash{\SetFigFont{12}{14.4}{\rmdefault}{\mddefault}{\updefault}\special{ps: gsave 0 0 0 setrgbcolor}K\special{ps: grestore}}}}
\put(3241,209){\circle{90}}
\put(2341,209){\circle{90}}
\put(2791,-286){\circle{90}}
\put(1891,209){\circle{90}}
\put(1081,209){\circle{90}}
\put(631,-286){\circle{90}}
\put(3601,209){\circle{90}}
\put(4411,209){\circle{90}}
\put(4861,-286){\circle{90}}
\end{picture}  
\end{array} 
\right) + k \leftrightarrow p 
\end{eqnarray}
Note that at all branch points there is momentum conservation: the
momentum flowing in must be equal to the momentum flowing out.  Black
dots represents summation, open dots integration.

All the diagrams consist of a few elementary building blocks, e.g.
lines with one, two or more open dots and one black dot and lines
between two black dots. The integrals corresponding to these building
blocks can be performed in a straightforward manner.  With these
building blocks one can construct all the diagrams.  Finally, to
obtain the correlation function, one has to perform the
summation. This can be done either analytically or by explicit
summation.  Performing the summation analytically is only feasible
when the architecture is regular.

This diagrammatic representation of the correlation functions
presented, is similar to the diagrammatical techniques used in
particular in ref \cite{Alexander} (also \cite{Read}).
\end{appendix}

\begin{figure}
\scalebox{0.4}[0.4]{
\rotatebox{-90}{
\includegraphics*{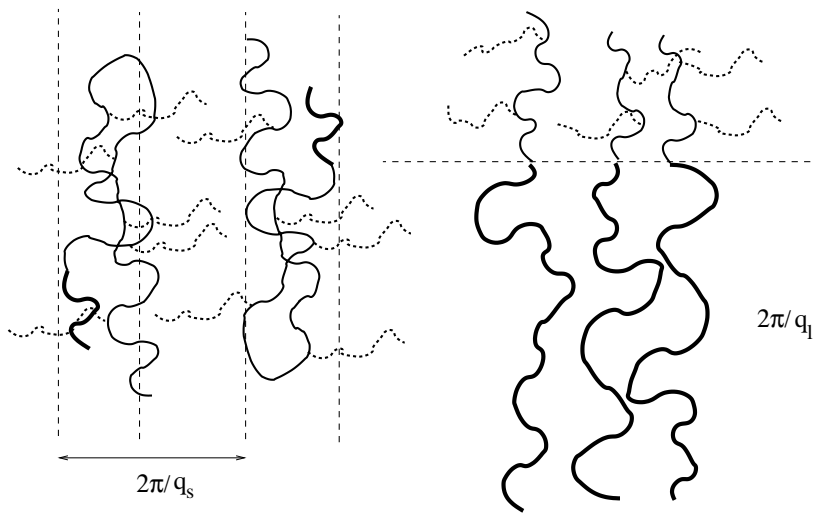}}}
\caption{
Illustration of the two ways in which the system can
microphase separate. 
}
\label{figII}
\end{figure}
\begin{figure}
\scalebox{0.5}[0.5]{
\rotatebox{-90}{
\includegraphics{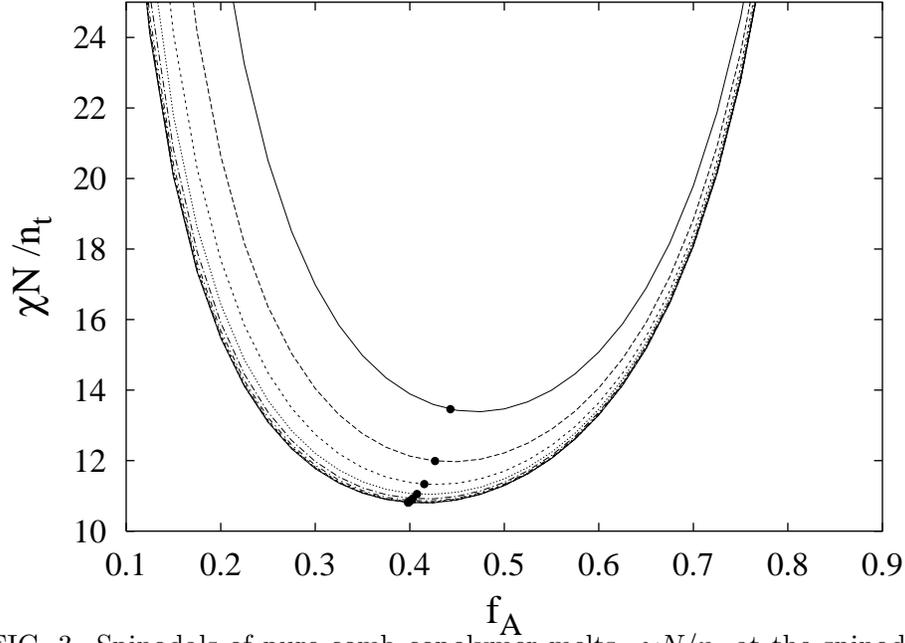}}}
\caption{Spinodals of pure comb copolymer melts. $\chi N/n_t$ at the
spinodal as a function of $f_A$ for the asymmetry parameter $t=\half$
and $\alpha=1$ (one side chain per branch point).  $(A_2B_1)_{n_t}$
Going from top to bottom the lines correspond to the number of branch
points $n_t=1,2,4,\cdots,1024$.  The dots correspond to the critical
points.}
\label{figIII}
\end{figure}
\begin{figure}
\scalebox{0.5}[0.5]{
\rotatebox{-90}{
\includegraphics{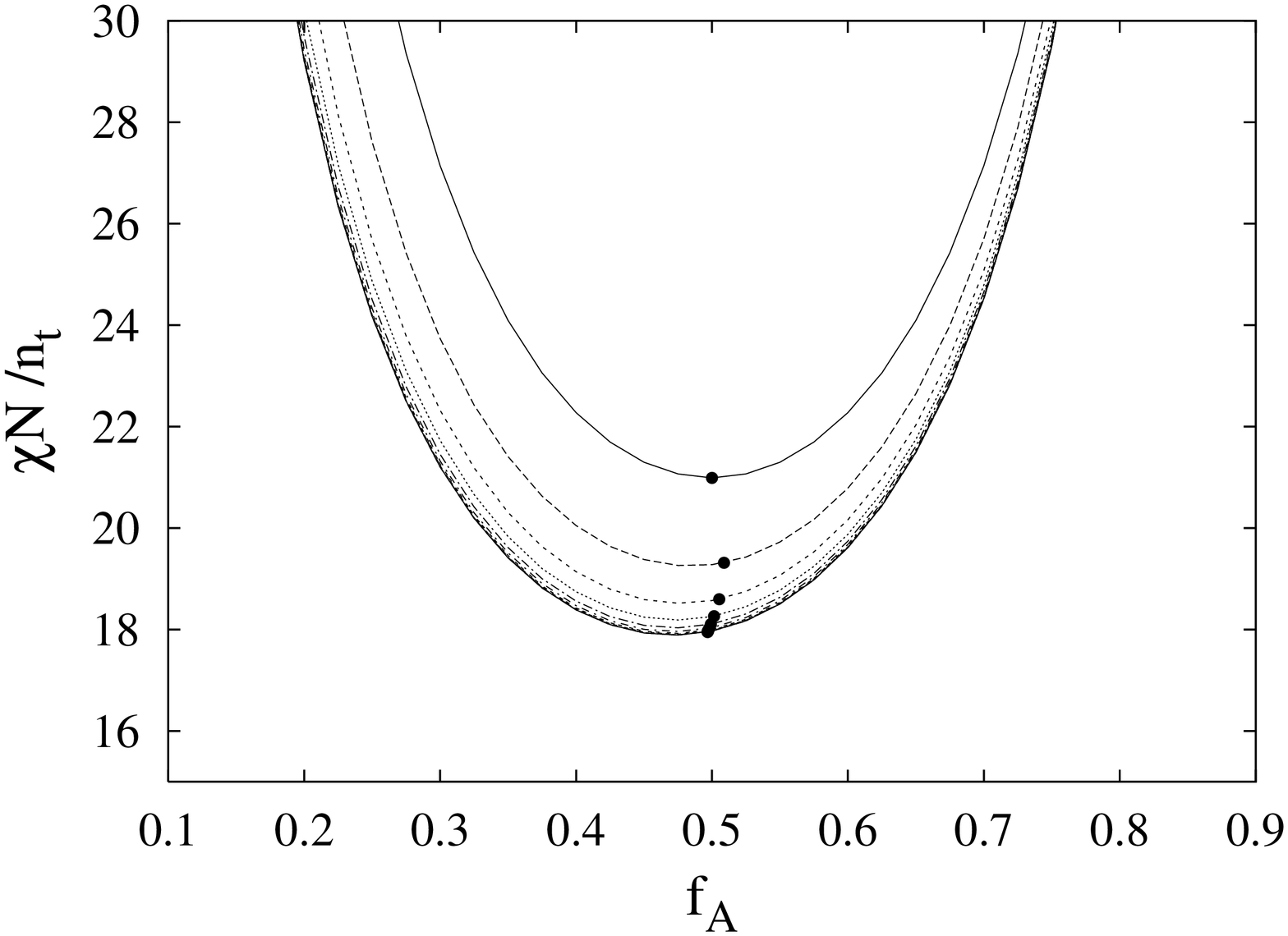}}}
\caption{Spinodals of pure comb copolymer melts. $\chi N/n_t$ at the
spinodal as a function of $f_A$ for the asymmetry parameter $t=\half$
and two side chains per branch point, $\alpha=2$.
$(A_2B_2)_{n_t}$. Going from top to bottom the lines correspond to the
number of branch points $n_t=1,2,4,\cdots,1024$. The dots correspond
to the critical points.}
\label{figIV}
\end{figure}
\begin{figure}
\scalebox{0.5}[0.5]{
\rotatebox{-90}{
\includegraphics{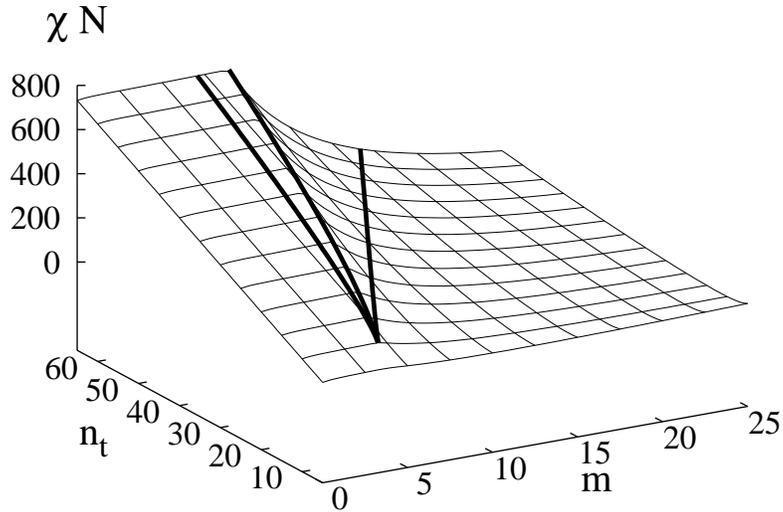}}}
\caption{Spinodal value $(\chi N)_s$ for the class of comb-coil
diblock copolymers defined in the text versus $(n_t,m)$. The "outer"
heavy lines on the $\chi N$ surface indicate the boundary between the
region where $\Gamma_2$ has one minimum and the region where
$\Gamma_2$ has two minima. Along the "inner" heavy line the two minima
have the same value; it separates ordering on a short length scale
from ordering on a long length scale. }
\label{figV}
\end{figure}
\begin{figure}
\scalebox{0.5}[0.5]{
\rotatebox{-90}{
\includegraphics{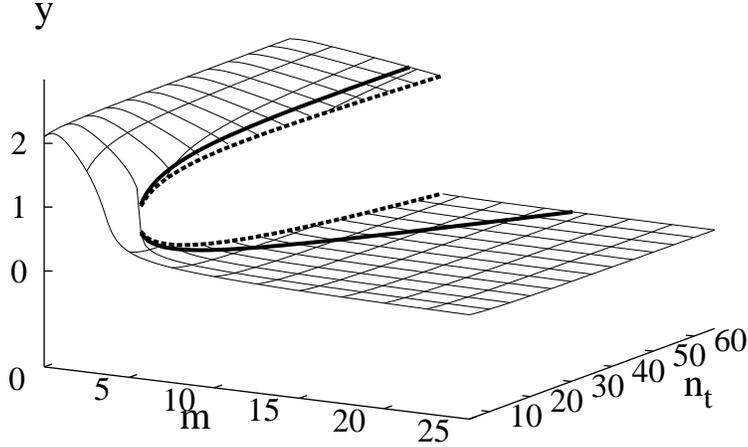}}}
\caption{Spinodal value $y$ versus $(n_t,m)$ for the class of
comb-coil diblock copolymers defined in the text., Here $y= {q^*}^2
R^2_g$. $R_g^2=a^2 d /6$ is the radius of gyration of a chain of
length $d$. $y$ is a dimensionless quantity corresponding to the
inverse length scale of microphase separation.  For small values of
$n_t$, below the bifurcation point (cf.\ref{figVII}), $\Gamma_2$ has
only one minimum and $y$ changes continuously as a function of $m$.
For large values of $n_t$, above the bifurcation point, $y$ changes
discontinuously from large $y$ (small length scale) to small $y$
(large length scale). Here $\Gamma_2$ has two minima and the absolute
minimum 'jumps' from the small to the large length scale. The heavy solid
lines delineate the region in which $\Gamma_2$ has two minima. Along the  heavy
dashed lines both minima of $\Gamma_2$ have the same value and these lines
therefore correspond to a sudden change in length scale.  }
\label{figVI}
\end{figure}
\begin{figure}
\scalebox{0.5}[0.5]{
\rotatebox{-90}{
\includegraphics{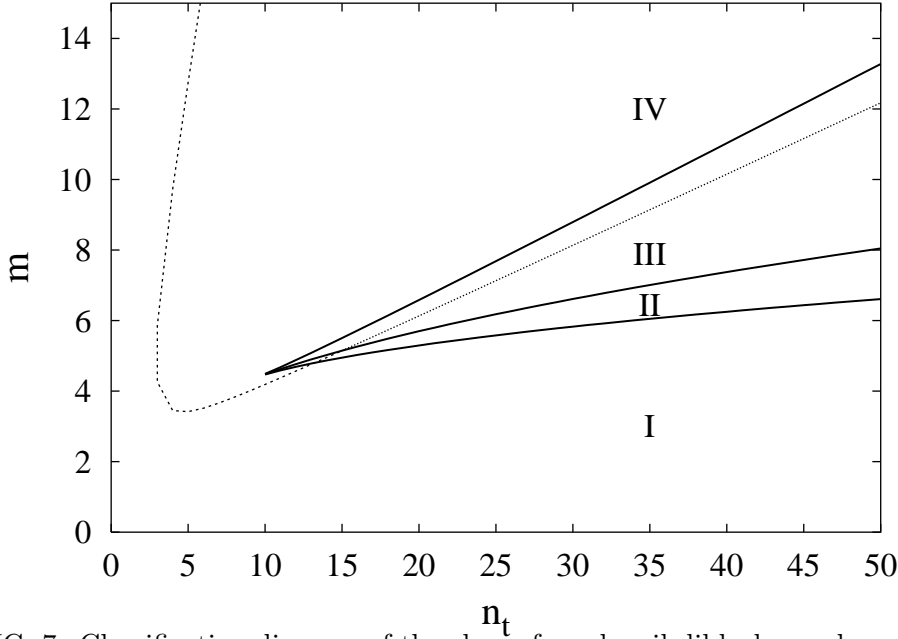}}}
\caption{Classification diagram of the class of comb-coil diblock
copolymer melts defined in the text. The solid lines correspond to the
projections of the boundary lines in figures \ref{figV} and
\ref{figVI} onto the $(n_t,m)$ plane. Region I: $\Gamma_2$ has one
minimum corresponding to the short length scale; Region II: $\Gamma_2$
has two minima, the absolute minimum corresponds to the short length
scale; Region III: $\Gamma_2$ has two minima, the absolute minimum
corresponds to the large length scale; Region IV: $\Gamma_2$ has one
minimum corresponding to the large length scale. The dashed and dotted
line presents the critical points. Inside region III, indicated with
the dotted part, the critical points correspond to the relative rather
than absolute minima of $\Gamma_2(q)$}
\label{figVII}
\end{figure}
\begin{figure}
\scalebox{0.5}[0.5]{
\rotatebox{-90}{
\includegraphics{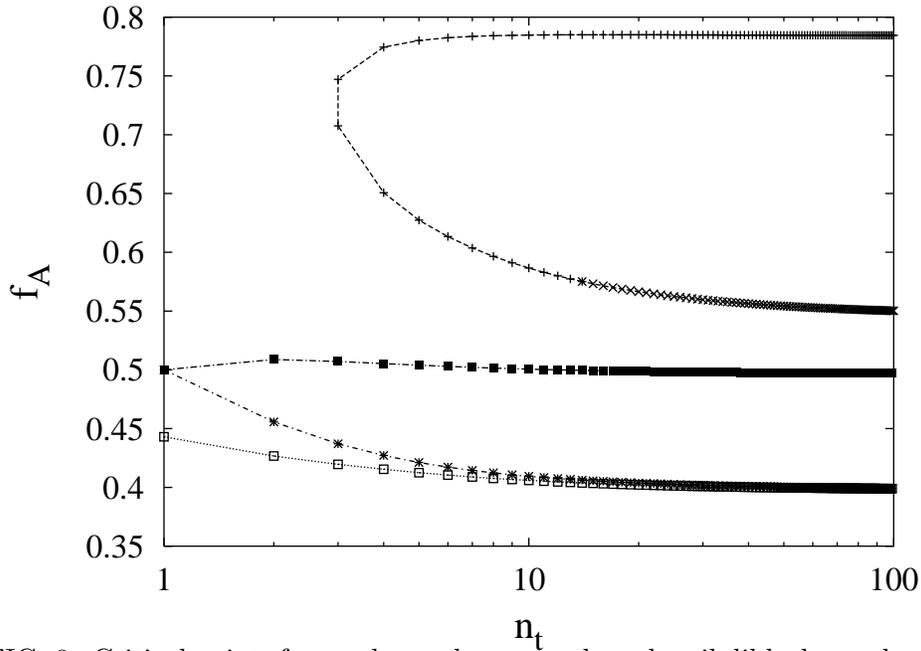}}}
\caption{Critical points for comb copolymers and comb-coil diblock
copolymers.  The lower three lines indicated with open boxes, stars
and filled boxes correspond to pure comb copolymers with
$(\alpha,t)=(1,\half),(1,0)$ and $(2,\half)$ respectively. The upper
curve corresponds to the comb-coil diblock copolymer.  The part of its
lower branch indicated with crosses ($n_t>15$) denote the 'critical points'
located in region III, see also Figure \ref{figVII} }
\label{figXII}
\end{figure}
\begin{figure}
\scalebox{0.5}[0.5]{
\rotatebox{-90}{
\includegraphics{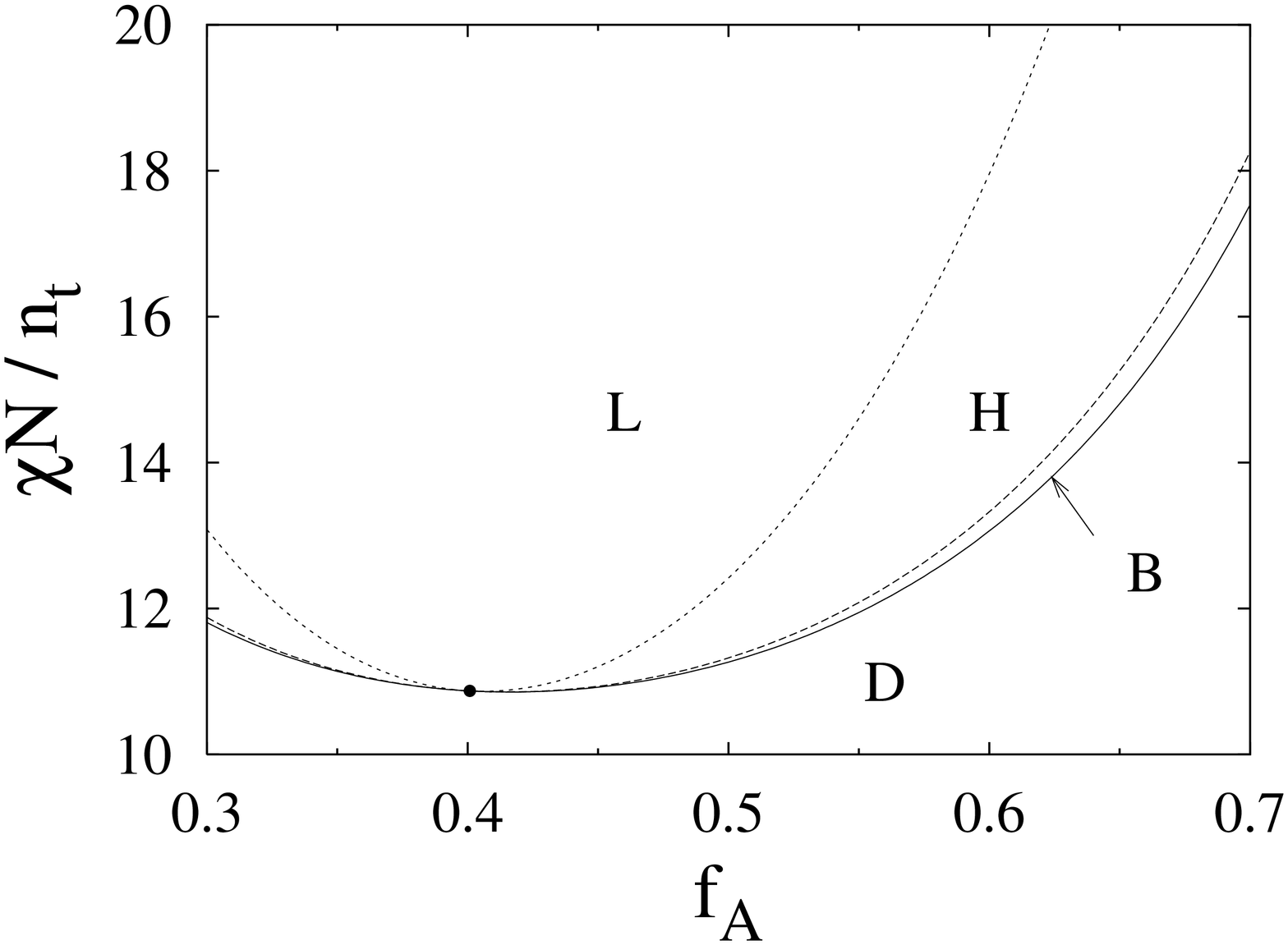}}}
\caption{Phase diagram of a pure comb copolymer melt with number of
branch points $n_t=30$, asymmetry parameter $t=\half$ and one side
chain per branch point, $\alpha=1$.  D = disordered, L = lamellar, H =
hexagonal and B = bcc. }
\label{figVIII}
\end{figure}
\begin{figure}
\scalebox{0.5}[0.5]{
\rotatebox{-90}{
\includegraphics{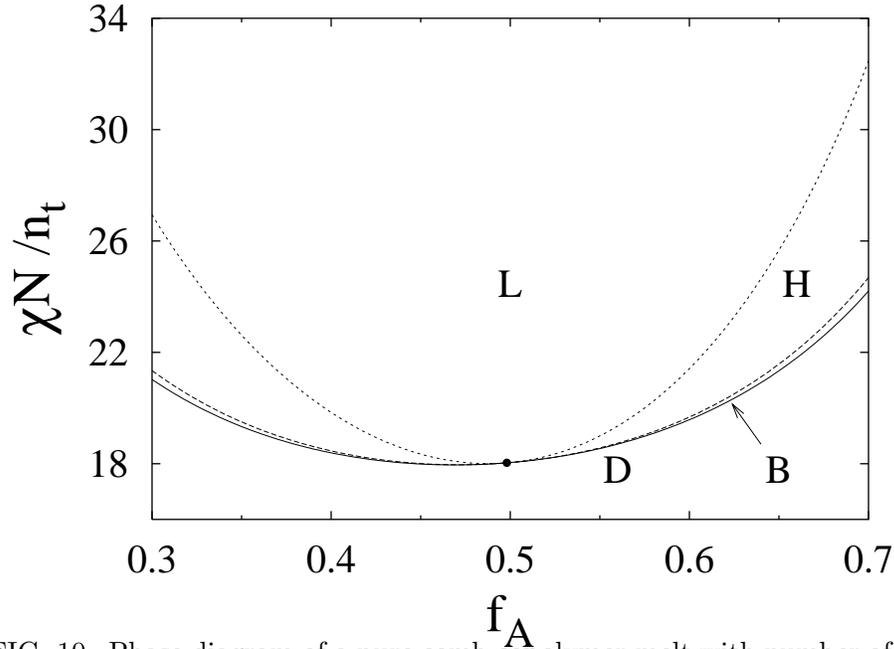}}}
\caption{Phase diagram of a pure comb copolymer melt with number of
branch points $n_t=30$, asymmetry parameter $t=\half$ and two side
chains per branch point, $\alpha=2$, i.e., two side chains per branch
point.  D = disordered, L = lamellar, H = hexagonal and B = bcc. }
\label{figIX}
\end{figure}
\begin{figure}
\scalebox{0.5}[0.5]{
\rotatebox{-90}{
\includegraphics{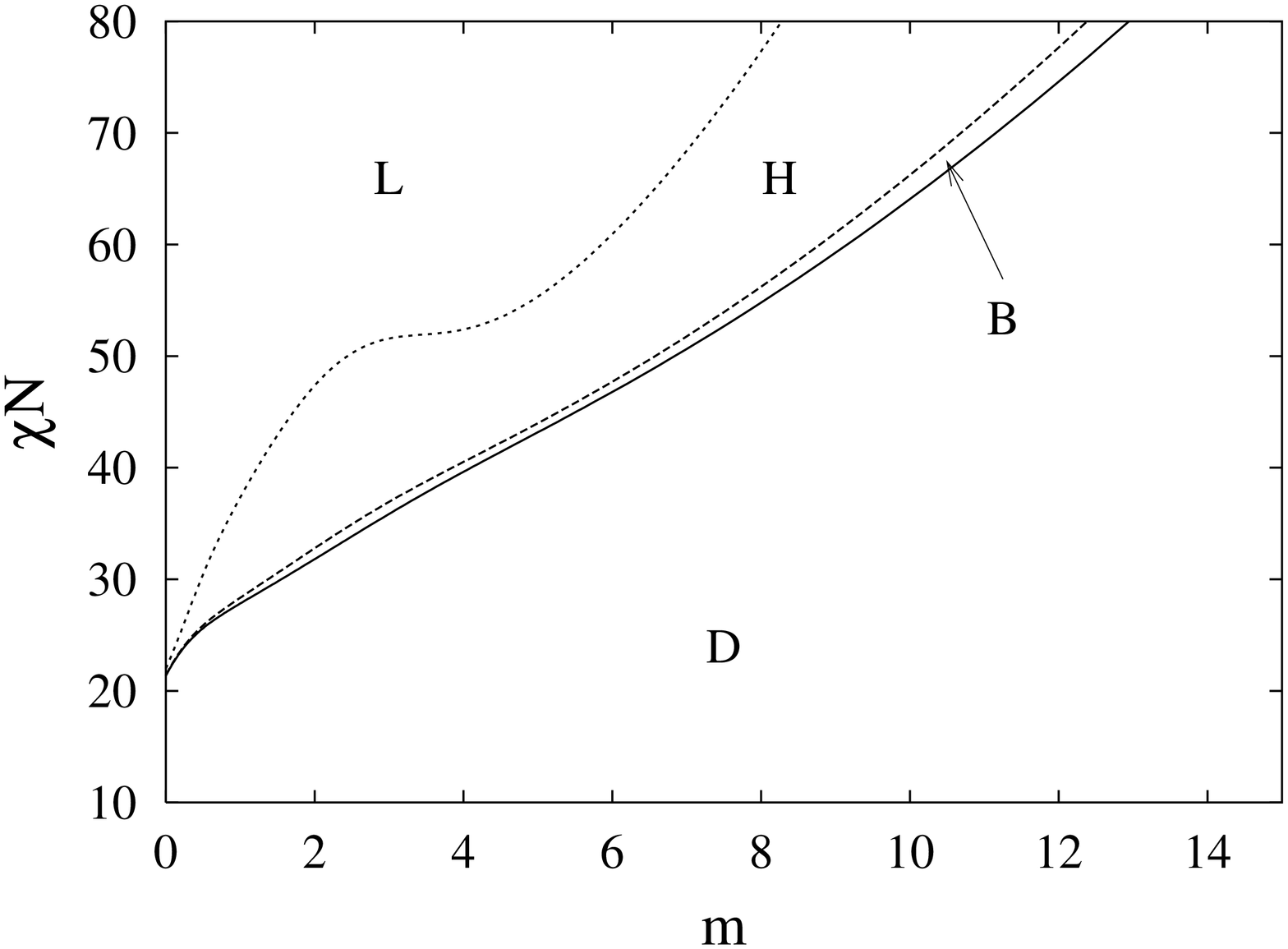}}}
\caption{Phase diagram for the class of comb-coil diblock copolymer
melts defined in the text. The figure corresponds to molecules with a
comb-block characterized by two branch points $n_t=2$, asymmetry
parameter $t=0$ and one side chain per branch point $\alpha=1$. Note
that $m$ corresponds to the length of the homopolymer block.(
A$_m$-{\it b}-(A-{\it g}-B)$_{2}$).  D = disordered, L = lamellar, H =
hexagonal and B = bcc.}
\label{figX}
\end{figure}
\begin{figure}
\scalebox{0.5}[0.5]{
\rotatebox{-90}{
\includegraphics{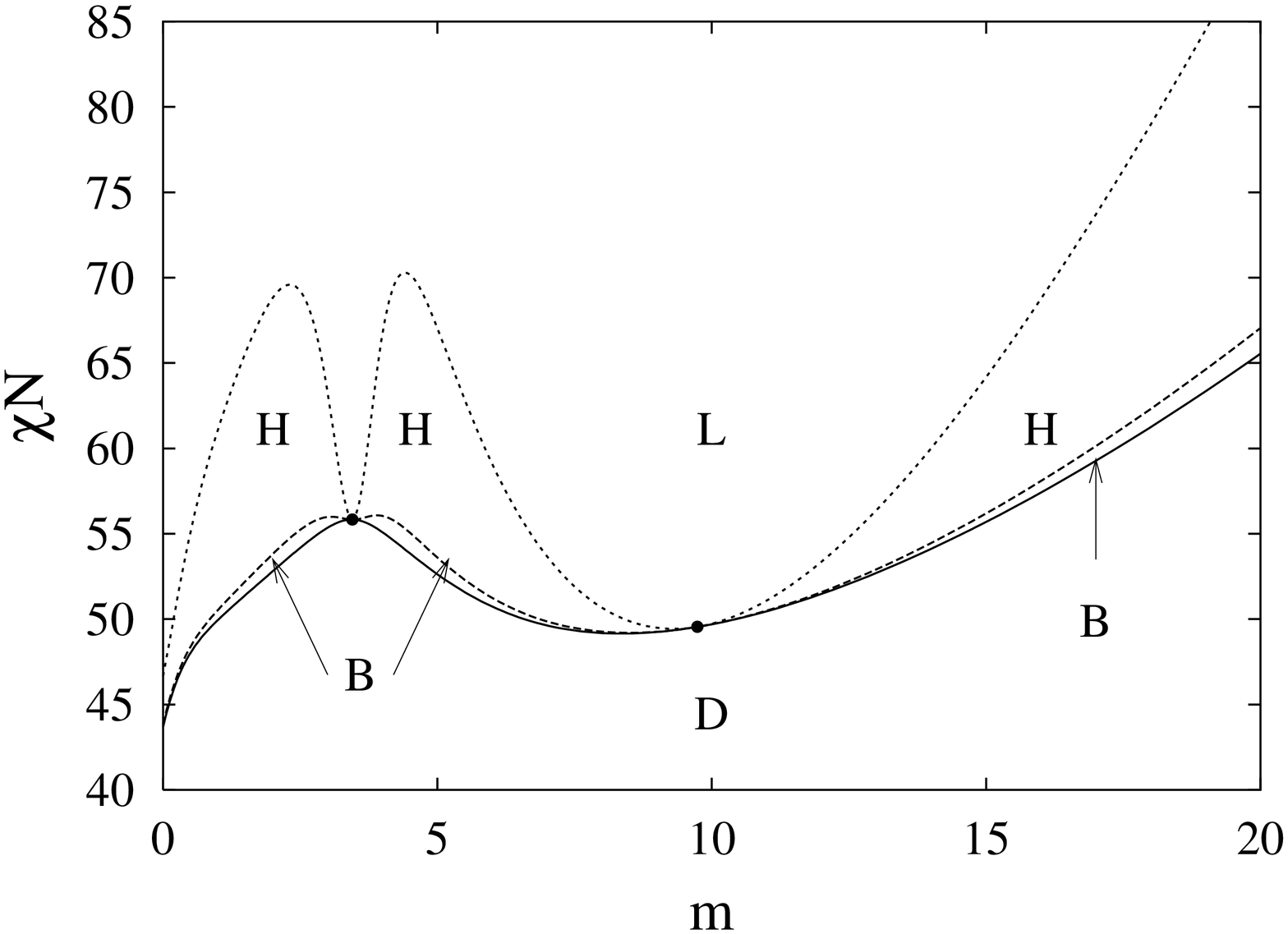}}}
\caption{Phase diagram for the class of comb-coil diblock copolymer
melts defined in the text. Figure corresponds to molecules with the
comb-block having four branch points $n_t=4$, asymmetry parameter
$t=0$ and one side chain per branch point $\alpha=1$. ( A$_m$-{\it
b}-(A-{\it g}-B)$_{4}$). Dots indicate critical points.  D =
disordered, L = lamellar, H = hexagonal and B = bcc.}
\label{figXI}
\end{figure}


\begin{thebibliography}{99}

\bibitem{Breiner} Breiner, U.; Krappe, U.; Abetz, V.; Stadler, R.
{\it Macromol. Chem. Phys.}  {\bf 1997}, {\it198}, 1051.

\bibitem{Werner} 
Werner, A.; Fredrickson, G. H. {\it J Polym Sci B: Polym Phys} 
{\bf 1997}, {\it35}, 849. 

\bibitem{Breiner2}
Breiner, U.; Krappe, U.; E.L. Thomas, E. L.; Stadler, R.
{\it Macromolecules} {\bf 1998}, {\it31}, 135. 

\bibitem{Helsinki1}
Ruokolainen, J.; M\"akinen, R.; Torkkeli, M.;M\"akel\"a, T.; Serimaa, R.; ten 
Brinke, G.; Ikkala, O. {\it Science} {\bf 1998}, {\it280}, 557.

\bibitem{Helsinki2}
Ruokolainen, J.; ten Brinke, G.; Ikkala, O. 
{\it Adv. Mater.} {\bf 1999}, {\it11}, 777.

\bibitem{Helsinki3} 
Ruokolainen, J.; Saariaho, M.; Ikkala, O.; ten Brinke, G.; Thomas, E. L.;
Torkkeli, M.; Serimaa, R.
{\it Macromolecules}  {\bf 1999}, {\it32}, 1152. 

\bibitem{Ott}
Ott, H.; Abetz, V; Altst\"adt, V. 
{\it Macromolecules} {\bf 2001}, {\it34}, in press.

\bibitem{rikkert}
Nap, R. J.; Kok, C.; ten Brinke, G.; Kuchanov, S.I. 
{\it Eur. Phys. J. E.} accepted.

\bibitem{Beyer}
Beyer, F.L.; Gido, S. P.; B\"uschl, C.; Iatrou, H.; Uhrig, D.;
 Mays, J.W.; Chang, M.Y.; Garetz, B.A.; Balsara, N. P.; 
Beck Tan, N.; Hadjichristidis, N.
{\it Macromolecules} {\bf 2000}, {\it33}, 2039. 

\bibitem{Tsoukatos}
Tsoukatos, T.; Pispas, S.; Hadjichristidis, N. 
{\it Macromolecules}, {\bf 2000}, {\it33}, 9504. 

\bibitem{Balasz2}
Shinozaki, A.; Jasnow, D.; Balazs, A.
{\it Macromolecules} {\bf 1994}, {\it27}, 2496.

\bibitem{Balasz1}
Foster, D. P.; Jasnow, D.; Balazs, A. 
{\it Macromolecules} {\bf 1995}, {\it28}, 3450.

\bibitem{Benoit}
Benoit, H.; Hadziioannou, G. 
{\it Macromolecules} {\bf 1988}, {\it21}, 1449.

\bibitem{Dobrynin}
Dobrynin, A. V.; Erukhimovich, I. Ya. 
{\it Macromolecules} {\bf 1993}, {\it 26}, 276.

\bibitem{delaCruz}
Olvera de la Cruz, M.; Sanchez, I. C. 
{\it Macromolecules} {\bf 1986}, {\it 19}, 2501.

\bibitem{Edwards}
Doi, M.; Edwards, S. F.
{\it The Theory of Polymer Dynamics}, Oxford University Press, Oxford, 1986

\bibitem{Hong}
Hong, K. M.; Noolandi, J, {\it Macromolecules} {\bf 1981}, {\it 14}, 727.

\bibitem{Matsen0}
Matsen, M. W.; Schick, M. 
{\it Phys. Rev. Lett.} {\bf 1994}, {\it 72}, 2660. 

\bibitem{Leibler} 
Leibler, L. {\it Macromolecules} {\bf 1980}, {\it13}, 1602.

\bibitem{Vilgis}
Holyst, R.; Vilgis, T. A.
{\it Macromol. Theory and Sim.} {\bf 1996}, {\it 5}, 573, cond-mat/9603063.

\bibitem{FML} 
Fredrickson, G.H.; Milner, S.T.; Leibler, L.
{\it Macromolecules} {\bf 1992}, {\it 25}, 6341.

\bibitem{Slot}
Slot, J. J. M.; Angerman, H. J.; ten Brinke, G.
{\it J. Chem. Phys.} {\bf 1998}, {\it 109}, 8677. 

\bibitem{Mayes}
Mayes, A. M.; Olvera de la Cruz,  M.
{\it J. Chem. Phys.} {\bf 1989}, {\it 91}, 7228.

\bibitem{delaCruz2}
Olvera de la Cruz, M. {\it Phys. Rev. Lett.} {\bf 1991},{\it 67}, 85.

\bibitem{Jones}
Jones, J. L.; Olvera de la Cruz, M. 
{\it J. Chem. Phys.} {\bf 1994}, {\it 100}, 5272.

\bibitem{Hamley1}
Hamley, I. W.; Bates, F. S. {\it J. Chem. Phys.} {\bf 1994}, {\it100}, 6813.

\bibitem{Erukhimovich}
Erukhimovich, I. Ya. {JETP Lett.} {\bf 1996}, {\it 63}, 459.

\bibitem{Milner2}
Milner, S. T.; Omsted, P. D. {\it J. Phys. II France } {\bf 1997} {\it 7}, 249.
  
\bibitem{Aksimentiev}
Aksimentiev, A.; Holyst, R. 
{\it J. Chem. Phys.} {\bf 1999}, {\it 111}, 2329. 

\bibitem{Alexander}
Morozov, A. N.; Fraaije, J. G. E. M.  
{\it J. Chem. Phys.} {\bf 2001}, {\it 114}, 2452, cond-mat/0008246.  

\bibitem{Semenov}
Semenov, A. N.; Likhtman, A. E. 
{\it Macromolecules} {\bf 1998}, {\it31}, 9058. 

\bibitem{Read}
Read, D. J.  {\it Macromolecules} {\bf 1998}, {\it 31}, 899.

\end{thebibliography}
\end{document}